\documentclass[english,spanish,american]{article}
\usepackage[T1]{fontenc}
\usepackage{textcomp}
\usepackage[utf8]{inputenc}
\usepackage{geometry}
\usepackage{babel}
\addto\shorthandsspanish{\spanishdeactivate{~<>}}
\deactivatequoting

\usepackage{varwidth}
\usepackage{amsmath}
\usepackage{amssymb}
\usepackage{graphicx}
\usepackage[]
 {hyperref}

\makeatletter

\providecommand{\tabularnewline}{\\}
\newenvironment{cellvarwidth}[1][t]
    {\begin{varwidth}[#1]{\linewidth}}
    {\@finalstrut\@arstrutbox\end{varwidth}}

\usepackage{babel}

\addto\shorthandsspanish{\spanishdeactivate{~<>}}
\deactivatequoting

\makeatother

\begin{document}
\title{Lepton Flavor Violating Higgs Decays in a Minimal Doublet Left-Right
Symmetric Model with an Inverse See-Saw}
\author{M. Zeleny-Mora$^{1}$\thanks{moiseszeleny@gmail.com}, R. Gaitán-Lozano$^{1}$\thanks{regaitan@gmail.com},
R. Martinez$^{2}$\thanks{remartinezm@unal.edu.co}\\
 {\small$^{1}$Departamento de Física, FES-Cuautitlán, UNAM, C.P.
54770, Estado de México, México.}\\
{\small$^{2}$Departamento de Física, Universidad Nacional de Colombia,
K. 45 No. 26-85, Bogotá, Colombia.}}
\maketitle
\begin{abstract}
In this study, we analyze the lepton flavor violation (LFV) decays
Swithin the framework of the Doublet Left-Right Symmetric model (DLRSM),
based on the gauge group $SU\left(2\right)_{L}\otimes SU\left(2\right)_{R}\otimes U\left(1\right)_{B-L}$.
The model features an extended gauge and scalar sector, including
a bidoublet and two doublets which induce new charged currents interactions.
Spontaneous Symmetry Breaking (SSB) occurs in two stages, introducing
a new scale associated with the vacuum expectation value (VEV) of
the right-handed doublet $v_{R}$ assumed to lie above the electroweak
scale. Neutrino masses are generated via the inverse seesaw mechanism,
allowing sizable mixing between active and sterile neutrinos. We diagonalize
the full neutrino mass matrix and express the mixing in terms of physical
parameters. We compute the branching ratios for LFV Higgs decays as
functions of the heavy neutrino mass scale. Our numerical analysis
incorporates current experimental bounds and projected sensitivities,
highlighting viable regions of parameter space where LFV signals could
be observed at future colliders.
\end{abstract}

\section{Introduction }

The Standard Model (SM) of particle physics has been remarkably successful
in describing fundamental interactions at the electroweak scale. Nonetheless,
several open questions remain, including the origin of neutrino masses,
the nature of parity violation, and the possibility of lepton flavor
violation (LFV). The Left-Right Symmetric Model (LRSM) offers a compelling
framework to address these issues by extending the SM gauge group
to 
\[
SU\left(3\right)_{C}\otimes SU\left(2\right)_{L}\otimes SU\left(2\right)_{R}\otimes U\left(1\right)_{B-L},
\]

restoring left-right symmetry at higher energies\cite{Pati:1974yy,Mohapatra:1974gc,Mohapatra:1974hk,Senjanovic:1975rk}.
In contrast to the canonical version based on triplet scalar fields,
the Doublet Left-Right Symmetric Model (DLRSM) introduces a scalar
sector consisting of a bidoublet $\Phi$ and two doublets $\chi_{L}$
and $\chi_{R}$, which simplifies the scalar potential\cite{SENJANOVIC1979334}.

Unlike the canonical LRSM, which relies on scalar triplets, the DLRSM
provides a more economical alternative, avoiding the presence of doubly
charged scalars, which are subject to stringent constraints from colliders
searches, flavor-changing neutral currents (FCNC), electric dipole
moments (EDMs) and precision electroweak measurements \cite{Dev:2016dja,Bertolini:2014sua}. 

However, in the DLRSM, Majorana masses for neutrinos are not automatically
and require additional mechanisms. One such mechanism is the inverse
seesaw (ISS) which predicts right-handed neutrinos at low scale \cite{Mohapatra:1986bd,Dev:2012sg}.
Neutrino oscillation experiments confirm that neutrinos have tiny
masses, and the most widely accepted explanation is the Type-I see-saw
mechanism. This introduces right-handed neutrinos $\nu_{R}$ and a
large Majorana mass $M_{R}$, which in the case of three right-handed
neutrinos, leads to a $6\times6$ neutrino mass matrix $\mathcal{M}_{\nu}$.
In the limit $|m_{D}|\ll|M_{R}|$, the light neutrino mass matrix
$\mathcal{M}_{\text{light}}$ is approximately
\[
\mathcal{M}_{\text{light}}\approx-m_{D}^{\top}M_{R}^{-1}m_{D}.
\]

where, $m_{D}$ is the Dirac mass matrix. Typically, this requires
$M_{R}\approx10^{14}$ GeV, a scale beyond current experimental reach. 

The ISS offers an alternative by introducing three pair of fermionic
singlets $\left(N_{R},S\right)$. In addition to $M_{R}$ a new small
Majorana mass matrix $\mu$ for the singlets $S$ is included. Assuming
the hierarchy $|\mu|\ll|m_{D}|\ll|M_{R}|$, the light neutrino mass
matrix becomes 
\[
\mathcal{M}_{\text{light}}\approx m_{D}^{\top}\left(M_{R}^{\top}\right)^{-1}\mu M_{R}^{-1}m_{D},
\]
allowing right-handed neutrinos to reside at the TeV scale, potentially
within reach of current collider experiments.

A notable consequence of the neutrino mass generation is LFV. In the
SM charged lepton sector of SM extended with neutrino masses LFV process
such as $\mu\to e\gamma$, $\mu\to3e$ and $\mu$-$e$ conversion
in nuclei are highly suppressed due to the smallness of neutrino masses
and the Glashow-Iliopoulos-Maiani (GIM) mechanism, making them effectively
unobservable in current experiments \cite{Cheng:1976uq}. In contrast,
LFV Higgs decays (LFVHD) offer a promising probe of flavor structure
in the scalar sector. These decays are directly linked to fermion
mass generation and may be observable at current or future colliders. 

Previous studies have explored, LFVHD in various extension of the
SM, including the Type-I and inverse seesaw mechanisms \cite{PhysRevD.71.035011,PhysRevD.91.015001,THAO2017159},
the 331 model \cite{HUE201637} and the 2HDM type III \cite{Vicente2019HiggsLF,Arroyo-Urena:2023vfh}
where sizable branching ratios are possible. Future lepton colliders
could be sensitives to interesting LFV violation signals \cite{PhysRevD.108.036002}.
Recent work has also examined the potential of high-energy muon collider
($\mu C$) to probe LFV process like $h\to\mu\tau$ and $\mu\to e\gamma$
\cite{Asadi:2025dii}. Current and projected experimental bound are
summarized in Table \ref{tab:LFV_experimental_bounds} for $h\to\mu\tau$
and $\mu\to e\gamma$.

\begin{table}
\begin{centering}
\begin{tabular}{|c|c|c|}
\hline 
Observable & Curren Limit & Projected Limits\tabularnewline
\hline 
$\mathcal{BR}(\mu\to e\gamma)$ & $1.5\times10^{-13}$ MEG-II \cite{MEGII:2025gzr} & $6\times10^{-14}$MEG-II \cite{MEGII:2025gzr}\tabularnewline
\hline 
$\mathcal{BR}(\tau\to e\gamma)$ & $3.3\times10^{-8}$ BaBar \cite{BaBar:2009hkt} & $3\times10^{-9}$ Belle II\cite{Belle-II:2018jsg}\tabularnewline
\hline 
$\mathcal{BR}(\tau\to\mu\gamma)$ & $4.2\times10^{-8}$ Belle \cite{Belle:2021ysv} & $1.0\times10^{-9}$ Belle II \cite{Belle-II:2018jsg}\tabularnewline
\hline 
\hline 
$\mathcal{BR}(h\to\mu\tau)$ & $1.8\times10^{-3}$ATLAS-2023\cite{ATLAS:2023mvd} & \begin{cellvarwidth}[t]
\centering
$5\times10^{-4}$ HL-LHC\\
$7.7\times10^{-5}$ $\mu C$ \cite{Asadi:2025dii}
\end{cellvarwidth}\tabularnewline
\hline 
$\mathcal{BR}(h\to\mu e)$ & $4.4\times10^{-5}$ CMS-2023 \cite{CMS:2023pte} & \begin{cellvarwidth}[t]
\centering
$1\times10^{-5}$ HL-LHC\\
$9.9\times10^{-6}$ $\mu C$ \cite{Asadi:2025dii}
\end{cellvarwidth}\tabularnewline
\hline 
$\mathcal{BR}(h\to\tau e)$ & $2\times10^{-3}$ ATLAS-2023\cite{ATLAS:2023mvd} & \begin{cellvarwidth}[t]
\centering
$5\times10^{-4}$ HL-LHC

$8.4\times10^{-5}$ $\mu C$ \cite{Asadi:2025dii}
\end{cellvarwidth}\tabularnewline
\hline 
\end{tabular}
\par\end{centering}
\caption{Current and projected upper bounds for LFV decays.}\label{tab:LFV_experimental_bounds}
\end{table}

The paper is organized as follows, in Section \ref{sec:Model} we
review the DLRSM model, analyzing the gauge and scalar sectors. Following,
the ISS is analyzed and the interactions of Yukawa sector are derived
in Section \ref{sec:ISS}. In Section \ref{sec:LFVHD}, we study the
LFVHD at one loop. We proceed to do a numerical analysis of the parameter
space of the model and its impact over LFVHD in the Section \ref{sec:Numerical-Analysis}.
We conclude in Section \ref{sec:Conclusions}. Also, we add four Appendix
to provide diagonalization of neutral gauge boson in Appendix \ref{sec:Gauge_diagonalization}.
Feynman rules are given in Appendix \ref{sec:Gauge_diagonalization}
and one loop form factors for LFV Higgs decays are given in Appendix
\ref{sec:One-loop}. 

\section{The Doublet Left-Right Symmetric Model}\label{sec:Model}

This model is based on the gauge group $SU\left(2\right)_{L}\otimes SU\left(2\right)_{R}\otimes U\left(1\right)_{B-L}$,
augmented by a LR symmetry\cite{SENJANOVIC1979334}. In this model,
fermions come in LR symmetric doublet representations $Q_{L,R}=\left(u,d\right)_{L,R}^{\top}$
and $L_{L,R}=\left(\nu,\ell\right)_{L,R}^{\top}$. Under $\mathcal{P}$
the LR symmetry impose $\Psi_{L}\leftrightarrow\Psi_{R}$ with $\Psi=Q,L$,
and the quantum numbers are
\[
L_{iL}=\begin{pmatrix}\nu_{i}^{\prime}\\
\ell_{i}^{\prime}
\end{pmatrix}_{L}:\left(2,1,-1\right),\qquad L_{iR}=\begin{pmatrix}\nu_{i}^{\prime}\\
\ell_{i}^{\prime}
\end{pmatrix}_{R}:\left(1,2,-1\right)
\]
\[
Q_{iL}=\begin{pmatrix}u_{i}^{\prime}\\
d_{i}^{\prime}
\end{pmatrix}_{L}:\left(2,1,1/3\right),\qquad L_{iR}=\begin{pmatrix}r_{i}^{\prime}\\
d_{i}^{\prime}
\end{pmatrix}_{R}:\left(1,2,1/3\right).
\]
$i=1,2,3$ runs over fermion generations. Also, we will add three
fermionic singlets $S_{i}$. Under parity the fermions transforms
as follows
\begin{equation}
L_{L}\longleftrightarrow L_{R},\qquad S\longleftrightarrow S^{c}.\label{eq:LR_tranformstion_fermions}
\end{equation}
In this model, the electric charge of particles are related with the
eigenvalues of the generators of $SU(2)_{L,R}$ and $U\left(1\right)_{B-L}$
groups as follows
\[
Q=T_{3L}+T_{3R}+\frac{B-L}{2}.
\]

In addition to the most common gauge boson $\overrightarrow{W}_{L}^{\mu}$
and $B^{\mu}$, there are three new gauge bosons associated to $SU(2)_{R}$,
denoted as $\overrightarrow{W}_{R}^{\mu}$. Then, left and right fermion
doublets $\Psi_{L,R}$ have each one a covariant derivative given
by
\[
D_{\mu}\Psi_{L}=\left(\partial_{\mu}-ig_{L}\frac{\vec{\tau}}{2}\cdot\vec{W}_{L\mu}-ig^{\prime}\frac{Y}{2}B_{\mu}\right)\Psi_{L},
\]
\[
D_{\mu}\Psi_{R}=\left(\partial_{\mu}-ig_{R}\frac{\vec{\tau}}{2}\cdot\vec{W}_{R\mu}-ig^{\prime}\frac{Y}{2}B_{\mu}\right)\Psi_{R},
\]
where the hypercharge $Y$ is defined from the Gell-Mann Nishijima
relation $Q=T_{3L}+\frac{Y}{2}$. We assume $g_{L}=g_{R}$ which is
called the Manifest Left-Right Symmetry (MLRS). Then, we have a fermion
gauge interaction Lagrangian as follows
\[
\mathcal{L}_{F}=\sum_{\Psi=Q,L}\left(\bar{\Psi}_{L}\gamma^{\mu}D_{\mu}\Psi_{L}+\bar{\Psi}_{R}\gamma^{\mu}D_{\mu}\Psi_{R}\right).
\]

\subsection{Higgs sector}

The Higgs sector consist of one bidoublet $\Phi\left(2,2,0\right)$
containing the usual SM Higgs field, with the decomposition 
\[
\Phi=\left[\phi_{1},i\sigma_{2}\phi_{2}^{*}\right],\quad\phi_{i}=\begin{pmatrix}\phi_{i}^{0}\\
\phi_{i}^{-}
\end{pmatrix}\quad\text{with }i=1,2;\quad\tilde{\Phi}=\sigma_{2}\Phi^{*}\sigma_{2}.
\]
The Vacuum Expectation Value (VEV) of $\Phi$ can be written as 
\[
\left\langle \Phi\right\rangle =\text{diag}\left(k_{1},k_{2}\right).
\]
In addition this model have two doublets 
\[
\chi_{L,R}=\begin{pmatrix}\chi_{L,R}^{+}\\
\chi_{L,R}^{0}
\end{pmatrix}_{L,R}\qquad\left\langle \chi_{L,R}\right\rangle =\begin{pmatrix}0\\
v_{L,R}
\end{pmatrix};
\]
with the following quantum numbers 
\[
\begin{array}{ll}
\chi_{L}\left(2,1,1\right) & \chi_{R}\left(1,2,1\right)\end{array}
\]
Under parity, the scalar multiplets transform as 
\begin{equation}
\chi_{L}\longleftrightarrow\chi_{R},\qquad\Phi\longleftrightarrow\Phi^{\dagger}.\label{eq:LR_transformation_multiplets}
\end{equation}

In this context, the scalar potential is given by \cite{SENJANOVIC1979334}
\begin{equation}
\begin{aligned}V\left(\chi_{L},\chi_{R},\Phi\right)= & -\mu_{1}^{2}\operatorname{Tr}\Phi^{\dagger}\Phi+\lambda_{1}\left(\operatorname{Tr}\Phi^{\dagger}\Phi\right)^{2}+\lambda_{2}\operatorname{Tr}\Phi^{\dagger}\Phi\Phi^{\dagger}\Phi+\frac{1}{2}\lambda_{3}\left(\operatorname{Tr}\Phi^{\dagger}\tilde{\Phi}+\operatorname{Tr}\tilde{\Phi}^{\dagger}\Phi\right)^{2}\\
 & +\frac{1}{2}\lambda_{4}\left(\operatorname{Tr}\Phi^{\dagger}\tilde{\Phi}-\operatorname{Tr}\tilde{\Phi}^{\dagger}\Phi\right)^{2}+\lambda_{5}\operatorname{Tr}\Phi^{\dagger}\Phi\tilde{\Phi}^{\dagger}\tilde{\Phi}+\frac{1}{2}\lambda_{6}\left[\operatorname{Tr}\Phi^{\dagger}\tilde{\Phi}\Phi^{\dagger}\tilde{\Phi}+\text{ h.c. }\right]\\
 & -\mu_{2}^{2}\left(\chi_{\mathrm{L}}^{\dagger}\chi_{\mathrm{L}}+\chi_{\mathrm{R}}^{\dagger}\chi_{\mathrm{R}}\right)+\rho_{1}\left(\left(\chi_{\mathrm{L}}^{\dagger}\chi_{\mathrm{L}}\right)^{2}+\left(\chi_{\mathrm{R}}^{\dagger}\chi_{\mathrm{R}}\right)^{2}\right)+\rho_{2}\chi_{\mathrm{L}}^{\dagger}\chi_{\mathrm{L}}\chi_{\mathrm{R}}^{\dagger}\chi_{\mathrm{R}}\\
 & +\alpha_{1}\operatorname{Tr}\Phi^{\dagger}\Phi\left(\chi_{\mathrm{L}}^{\dagger}\chi_{\mathrm{L}}+\chi_{\mathrm{R}}^{\dagger}\chi_{\mathrm{R}}\right)+\alpha_{2}\left(\chi_{\mathrm{L}}^{\dagger}\Phi\Phi^{\dagger}\chi_{\mathrm{L}}+\chi_{\mathrm{R}}^{\dagger}\Phi^{\dagger}\Phi\chi_{\mathrm{R}}\right)\\
 & +\alpha_{3}\left(\chi_{\mathrm{L}}^{\dagger}\tilde{\Phi}\tilde{\Phi}^{\dagger}\chi_{\mathrm{L}}+\chi_{\mathrm{R}}^{\dagger}\tilde{\Phi}^{\dagger}\tilde{\Phi}\chi_{R}\right).
\end{aligned}
\label{eq:Scalar_potential}
\end{equation}
The parameters $\mu_{1,2}^{2}$, $\lambda_{1,2,3,4,5,6},$$\rho_{1,2}$,
and $\alpha_{1,2,3}$ are all real. We consider the case where there
is no explicit spontaneous CP violation. The RH doublet $\chi_{R}$
is responsible for the breaking of $G_{LR}$ down to the SM gauge
symmetry $SU\left(2\right)_{L}\otimes U\left(1\right)_{Y}$, and its
non-vanishing VEV $v_{R}$ gives masses to the new heavy gauge boson
$W_{R}$ and $Z_{R}$ and the RH neutrinos $\nu_{R}$ . The bidoublet
$\Phi$ is responsible of the mass matrices of the ordinary fermions
in the SM after the Spontaneous Symmetry Breaking (SSB).

The neutral fields $\phi_{1,2}^{0}$, $\chi_{R,L}^{0}$ can be decomposed
in terms of real and imaginary part, ($\phi=\phi^{r}+i\phi^{i}$ with
$\phi=\phi_{1,2}^{0},\chi_{R,L}^{0}$). As a consequence from the
tadpole conditions 
\[
\frac{\partial V}{\partial\phi_{1}^{0r}}=\frac{\partial V}{\partial\phi_{2}^{0r}}=\frac{\partial V}{\partial\delta_{R}^{0r}}=\frac{\partial V}{\partial\delta_{L}^{0r}}=0
\]
we obtain the following equations 
\begin{equation}
\begin{aligned}\begin{aligned}\frac{\partial V}{\partial k_{1}}= & {\displaystyle 2k_{1}\left(-\mu_{1}^{2}+2k_{1}^{2}\left(\lambda_{1}+\lambda_{2}\right)+2k_{2}^{2}\left(\lambda_{1}+4\lambda_{3}+\lambda_{5}+\lambda_{6}\right)+v_{L}^{2}\left(\alpha_{1}+\alpha_{3}\right)+v_{R}^{2}\left(\alpha_{1}+\alpha_{3}\right)\right),}\\
\frac{\partial V}{\partial k_{2}}= & \ensuremath{{\displaystyle \ensuremath{{\displaystyle 2k_{2}\left(-\mu_{1}^{2}+2k_{1}^{2}\left(\lambda_{1}+4\lambda_{3}+\lambda_{5}+\lambda_{6}\right)+2k_{2}^{2}\left(\lambda_{1}+\lambda_{2}\right)+v_{L}^{2}\left(\alpha_{1}+\alpha_{2}\right)+v_{R}^{2}\left(\alpha_{1}+\alpha_{2}\right)\right)}},}}\\
\frac{\partial V}{\partial v_{L}}= & \ensuremath{{\displaystyle 2v_{L}\left(-\mu_{2}^{2}+2\rho_{1}v_{L}^{2}+\rho_{2}v_{R}^{2}+k_{1}^{2}\left(\alpha_{1}+\alpha_{3}\right)+k_{2}^{2}\left(\alpha_{1}+\alpha_{2}\right)\right)}},\\
\frac{\partial V}{\partial v_{R}}= & {\displaystyle 2v_{R}\left(-\mu_{2}^{2}+2\rho_{1}v_{R}^{2}+\rho_{2}v_{L}^{2}+k_{1}^{2}\left(\alpha_{1}+\alpha_{3}\right)+k_{2}^{2}\left(\alpha_{1}+\alpha_{2}\right)\right)}.
\end{aligned}
\end{aligned}
\label{eq:tadpole_eqs}
\end{equation}
In the case of $v_{L}=k_{2}=0$, from first and fourth tadpole conditions
\eqref{eq:tadpole_eqs}, we obtain $\mu_{1}^{2}$ and $\mu_{2}^{2}$,
as follows
\begin{align*}
\mu_{1}^{2}= & \ensuremath{{\displaystyle 2k_{1}^{2}\left(\lambda_{1}+\lambda_{2}\right)+v_{R}^{2}\left(\alpha_{1}+\alpha_{3}\right),}}\\
\mu_{2}^{2}= & \ensuremath{{\displaystyle 2\rho_{1}v_{R}^{2}+k_{1}^{2}\left(\alpha_{1}+\alpha_{3}\right)}}.
\end{align*}
The charged scalars, in the base $\left(\phi_{2}^{+},\chi_{\mathrm{L}}^{+},\phi_{1}^{+},\chi_{R}^{+}\right)$,
mass matrix is given by
\[
M_{+}^{2}=\begin{pmatrix}0 & 0 & 0 & 0\\
0 & k_{1}^{2}\left(\alpha_{2}-\alpha_{3}\right)+v_{R}^{2}\left(\rho_{2}-2\rho_{1}\right) & 0 & 0\\
0 & 0 & v_{R}^{2}\left(\alpha_{2}-\alpha_{3}\right) & k_{1}v_{R}\left(\alpha_{2}-\alpha_{3}\right)\\
0 & 0 & k_{1}v_{R}\left(\alpha_{2}-\alpha_{3}\right) & k_{1}^{2}\left(\alpha_{2}-\alpha_{3}\right)
\end{pmatrix},
\]
where two would be Goldstone boson emerge $G_{L,R}^{\pm}$ and two
charged scalars get mass as follows
\begin{align}
m_{H_{L}^{\pm}}^{2}= & k_{1}^{2}\left(\alpha_{2}-\alpha_{3}\right)+v_{R}^{2}\left(\rho_{2}-2\rho_{1}\right),\nonumber \\
m_{H_{R}^{\pm}}^{2}= & \left(\alpha_{2}-\alpha_{3}\right)\left(k_{1}^{2}+v_{R}^{2}\right).\label{eq:mHRpm}
\end{align}
In the limit $v_{R}\gg k_{1}$, we have
\begin{align}
\phi_{2}^{\pm}\approx & G_{L}^{\pm},\nonumber \\
\chi_{L}^{\pm}\approx & H_{L}^{\pm},\nonumber \\
\chi_{R}^{\pm}\approx & \frac{k_{1}H_{R}^{\pm}}{v_{R}}+G_{R}^{\pm},\nonumber \\
\phi_{1}^{\pm}\approx & {\displaystyle -\frac{k_{1}G_{R}^{\pm}}{v_{R}}+H_{R}^{\pm}}.\label{eq:Charged_scalars_mix}
\end{align}
In addition, two pseudo scalars $A_{1,2}^{0}$ obtain mass after the
SSB given by
\begin{align}
m_{A_{1}^{0}}^{2}= & 2v_{R}^{2}\left(\alpha_{2}-\alpha_{3}\right)+4k_{1}^{2}\left(-\lambda_{2}-4\lambda_{4}+\lambda_{5}-\lambda_{6}\right),\nonumber \\
m_{A_{2}^{0}}^{2}= & 2v_{R}^{2}\left(\rho_{2}-2\rho_{1}\right).\label{eq:Pseudoscalar_masses}
\end{align}
and two neutral would be Goldstone boson appear $G_{Z_{1}}$and $G_{Z_{2}}.$
In this sector we do not have mixings and the mass eigenstates are
\begin{align}
\chi_{L}^{0i}= & A_{2}^{0},\nonumber \\
\chi_{R}^{0i}= & G_{Z^{\prime}},\nonumber \\
\phi_{1}^{0i}= & G_{Z},\nonumber \\
\phi_{2}^{0i}= & A_{1}^{0}.\label{eq:neutral_pseudoscalar_mix}
\end{align}
Finally for neutral scalars, in the basis $\left(\phi_{2}^{0r},\chi_{L}^{0r},\phi_{1}^{0r},\chi_{R}^{0r}\right)$,
the mass matrix is given by
\[
M_{H}^{2}=\begin{pmatrix}2v_{R}^{2}\left(\alpha_{2}-\alpha_{3}\right)+4k_{1}^{2}\left(-\lambda_{2}+4\lambda_{3}+\lambda_{5}+\lambda_{6}\right) & 0 & 0 & 0\\
0 & 2v_{R}^{2}\left(\rho_{2}-2\rho_{1}\right) & 0 & 0\\
0 & 0 & 8k_{1}^{2}\left(\lambda_{1}+\lambda_{2}\right) & 4k_{1}v_{R}\left(\alpha_{1}+\alpha_{3}\right)\\
0 & 0 & 4k_{1}v_{R}\left(\alpha_{1}+\alpha_{3}\right) & 8\rho_{1}v_{R}^{2}
\end{pmatrix},
\]
As a consequence, we have four neutral Higgs scalars, where $\phi_{2}^{0r}$
and $\chi_{L}^{0r}$ are already physical fields. In the other hand,
$\phi_{1}^{0r}$ and $\chi_{R}^{0r}$ are mixed. Then, we have four
massive neutral scalars $H_{i}^{0}$, with $i=1,2,3,4$, with masses
given by
\begin{align}
m_{H_{1}^{0}}^{2}\approx & \ensuremath{{\displaystyle \left(8\left(\lambda_{1}+\lambda_{2}\right)-\frac{2\left(\alpha_{1}+\alpha_{3}\right)^{2}}{\rho_{1}}\right)k_{1}^{2},}}\nonumber \\
m_{H_{2}^{0}}^{2}\approx & \ensuremath{{\displaystyle 8\rho_{1}v_{R}^{2}+\frac{4}{\rho_{1}}\left(\alpha_{1}+\alpha_{3}\right)^{2}k_{1}^{2}},}\nonumber \\
m_{H_{3}^{0}}^{2}= & 2\left(\alpha_{2}-\alpha_{3}\right)v_{R}^{2}+4\left(-\lambda_{2}+4\lambda_{3}+\lambda_{5}+\lambda_{6}\right)k_{1}^{2},\nonumber \\
m_{H_{4}^{0}}^{2}= & 2\left(\rho_{2}-2\rho_{1}\right)v_{R}^{2}.\label{eq:Hi0_mass}
\end{align}
The mixing of the physics neutral scalars is given by
\begin{align}
\phi_{1}^{0r}\approx & \ensuremath{\left(\frac{k_{1}}{2\rho_{1}v_{R}}\left(\alpha_{1}+\alpha_{3}\right)\right)H_{2}^{0}+H_{1}^{0}},\nonumber \\
\chi_{R}^{0r}\approx & -\left(\frac{k_{1}}{2\rho_{1}v_{R}}\left(\alpha_{1}+\alpha_{3}\right)\right)H_{1}^{0}+H_{2}^{0},\nonumber \\
\phi_{2}^{0r}\approx & H_{3}^{0},\nonumber \\
\chi_{L}^{0r}\approx & H_{4}^{0}.\label{eq:neutral_scalar_mix}
\end{align}

In this context, there are 16 degrees of freedom which comes from
the 8 complex scalar fields in the multiplets $\chi_{L,R}$ and $\Phi$.
After SSB six massive bosons are produced $W_{L,R}^{\pm}$, $Z_{L,R}$,
six would-be Goldstone boson have been eaten $G_{L,R}^{\pm}$, $G_{Z_{1,2}}$,
leaving 10 degrees of freedom for the physical Higgs bosons. Four
scalars $H_{1,2,3,4}^{0}$, two pseudoscalars $A_{1,2}^{0}$ and four
charged scalars $H_{L,R}^{\pm}$ where $H_{1}^{0}$ is identified
with the SM Higgs $h^{SM}$.

\subsection{Kinetic Gauge sector}

In this case, the kinetic lagrangian for Higgs multiplets is given
by
\[
\mathcal{L}_{D}=\left(D_{\mu}\chi_{L}\right)^{\dagger}D_{\mu}\chi_{L}+\left(D_{\mu}\chi_{R}\right)^{\dagger}D_{\mu}\chi_{R}+\operatorname{Tr}\left[\left(D_{\mu}\Phi\right)^{\dagger}D_{\mu}\Phi\right],
\]
where the covariant derivatives are as follows ($g_{L}=g_{R}=g$)
\begin{align*}
D_{\mu}\chi_{L}= & \partial_{\mu}\chi_{L}-\frac{1}{2}ig\vec{\tau}\cdot\vec{W}_{L}\chi_{L}-ig_{B-L}B_{\mu},\\
D_{\mu}\chi_{R}= & \partial_{\mu}\chi_{R}-\frac{1}{2}ig\vec{\tau}\cdot\vec{W}_{R}\chi_{R}-ig_{B-L}B_{\mu},\\
D_{\mu}\Phi= & \partial_{\mu}\Phi-\frac{1}{2}ig\left(\vec{\tau}\cdot\vec{W}_{L}\Phi-\Phi\vec{\tau}\cdot\vec{W}_{R}\right).
\end{align*}
Similarly as the $W_{\mu}^{\pm}$ in the SM, we define, $W_{L,R\mu}^{\pm}\equiv\frac{1}{\sqrt{2}}\left(W_{L,R\mu}^{1}\mp iW_{L,R\mu}^{2}\right)$
and the mass matrix for charged gauge bosons is given by 
\begin{equation}
M_{W^{\pm}}^{2}=\begin{pmatrix}\frac{g^{2}\left(k_{1}^{2}+k_{2}^{2}+v_{R}^{2}\right)}{4} & -\frac{g^{2}k_{1}k_{2}}{2}\\
-\frac{g^{2}k_{1}k_{2}}{2} & \frac{g^{2}\left(k_{1}^{2}+k_{2}^{2}\right)}{4}
\end{pmatrix}\label{eq:Matrix_Wpm}
\end{equation}
the mixing angle $\xi$ of charged gauge bosons $W_{\mu}-W_{\mu}^{\prime}$
is given by
\begin{equation}
\tan\left|2\xi\right|=\frac{4k_{1}k_{2}}{v_{R}^{2}}\qquad\sin\xi\approx{\displaystyle \frac{2k_{1}k_{2}}{v_{R}^{2}}}.\label{eq:W_mixing_angle}
\end{equation}
 Then, the relation among $W_{\mu L,R}^{\pm}$ and the physical states
$W_{\mu}$ and $W_{\mu}^{\prime}$ is given by 
\begin{align}
W_{\mu}^{\pm}= & W_{\mu L}^{\pm}+\frac{2k_{1}k_{2}}{v_{R}^{2}}W_{\mu R}^{\pm},\nonumber \\
W_{\mu}^{\prime^{\pm}}= & W_{\mu R}^{\pm}-\frac{2k_{1}k_{2}}{v_{R}^{2}}W_{\mu L}^{\pm}.\label{eq:W_mixing}
\end{align}

where the mixing of the charged gauge bosons is tiny due to $k_{1},k_{2}\ll v_{R}$.
In the limit $k_{2}=0$, the mixing is null and the W gauge boson
mass are given by
\begin{align}
m_{W}^{2}\approx & \frac{g^{2}k_{1}^{2}}{4},\nonumber \\
m_{W^{\prime}}^{2}\approx & \frac{g^{2}v_{R}^{2}}{4}.\label{eq:Wprime_mass}
\end{align}

In the neutral gauge sector, in the basis $\left(W_{\mu L}^{3},W_{\mu R}^{3},B_{\mu}\right)$,
the mass matrix is given by
\begin{equation}
M_{Z}^{2}=\begin{pmatrix}\frac{g^{2}\left(k_{1}^{2}+k_{2}^{2}\right)}{4} & -\frac{g^{2}\left(k_{1}^{2}+k_{2}^{2}\right)}{4} & 0\\
-\frac{g^{2}\left(k_{1}^{2}+k_{2}^{2}\right)}{4} & \frac{g^{2}\left(k_{1}^{2}+k_{2}^{2}+v_{R}^{2}\right)}{4} & -\frac{gg_{B-L}v_{R}^{2}}{4}\\
0 & -\frac{gg_{B-L}v_{R}^{2}}{4} & \frac{g_{B-L}^{2}v_{R}^{2}}{4}
\end{pmatrix}\label{eq:MZ_mass_matrix}
\end{equation}
which is diagonalized in the Appendix \ref{sec:Gauge_diagonalization}
by the matrix $R_{Z}$in \eqref{eq:MZ_diagonal}. However, we consider
the limit of the $Z-Z^{\prime}$ mixing angle $\zeta$ \eqref{eq:Z_mixing_angle}is
null. As a consequence the weak gauge boson $W_{L}^{3},W_{R}^{3},B$,
are written in terms of the photon $A$ massless, $Z$ and $Z^{\prime}$gauge
boso, as follows 
\begin{align}
W_{\mu L}^{3}= & \ensuremath{{\displaystyle A_{\mu}\sin\theta_{W}-Z_{\mu}\cos\theta_{W}}},\nonumber \\
W_{\mu R}^{3}= & \ensuremath{{\displaystyle A_{\mu}\sin\theta_{W}+Z_{\mu}\sin\theta_{W}\tan\theta_{W}-Z_{\mu}^{\prime}\frac{\sqrt{\cos\left(2\theta_{W}\right)}}{\cos\theta_{W}}},}\nonumber \\
B_{\mu}= & \ensuremath{{\displaystyle A_{\mu}\sqrt{\cos\left(2\theta_{W}\right)}+Z_{\mu}\sqrt{\cos\left(2\theta_{W}\right)}\tan\theta_{W}+Z_{\mu}^{\prime}\tan\theta_{W}}}.\label{eq:Neutral_guage_mixings}
\end{align}
Finally, masses for the neutral gauge bosons in the limit of $k_{2}=0$
and $k_{1}\ll v_{R}$ are given by
\begin{align}
m_{Z}^{2}= & \frac{m_{W}^{2}}{\cos^{2}\theta_{W}}\nonumber \\
m_{Z^{\prime}}^{2}= & \ensuremath{m_{W^{\prime}}^{2}\frac{\cos^{2}\theta_{W}}{\cos\left(2\theta_{W}\right)}{\displaystyle -m_{W}^{2}\frac{\left(\tan\left(2\theta_{W}\right)+4\right)\tan^{2}\theta_{W}}{2}}.}\label{eq:ZZp_masses}
\end{align}

\section{The inverse see-saw }\label{sec:ISS}

In the lepton sector, the Dirac mass term is proportional to $\Phi$
and the Majorana mass term have contribution of both doublets $\chi_{L,R}$,
as follows \cite{Gu:2010xc,Brdar:2018sbk} 
\begin{equation}
\begin{aligned}-\mathcal{L}_{Y}= & \overline{L}_{iR}Y_{ij}\Phi^{\dagger}L_{jL}+\overline{L}_{iR}\tilde{Y}_{ij}\tilde{\Phi}^{\dagger}L_{jL}+\overline{S}_{i}Y_{ijL}\tilde{\chi}_{L}^{\dagger}L_{ijL}+\overline{S}_{i}^{c}Y_{ijR}\tilde{\chi}_{R}^{\dagger}L_{jR}+\frac{1}{2}\overline{S}_{i}^{c}\mu_{ij}S_{j}+\text{h.c.}\end{aligned}
\label{eq:LYukawa}
\end{equation}

where $Y$, $\tilde{Y}$, $Y_{L}$and $Y_{R}$ are $3\times3$ matrices
for Yukawa couplings and $\mu$ the Majorana mass matrix for fermionic
singlets $S_{i}$. In addition, $\tilde{X}\equiv i\sigma_{2}X^{*}$with
$X=\chi_{L},\chi_{R}$, $S^{c}=C\overline{S}^{\top}$and $\tilde{\Phi}=\sigma_{2}\Phi^{*}\sigma_{2}$,
denote charge conjugate fields of scalars and fermions. The transformations
under parity, following \eqref{eq:LR_tranformstion_fermions} and
\eqref{eq:LR_transformation_multiplets} impose the following relation
for Yukawa and Majorana matrices as follows
\[
Y_{L}=Y_{R},\qquad Y=Y^{\dagger},\qquad\tilde{Y}=\tilde{Y}^{\dagger},\qquad\mu=\mu^{\dagger},
\]

above the LR symmetry breaking scale. 

For charged leptons the mass matrix is given by
\begin{equation}
M_{\ell}=\frac{1}{\sqrt{2}}\left(k_{1}\tilde{Y}+k_{2}Y\right).\label{eq:Charged_lepton_mass_matrix}
\end{equation}
and the mass matrix is diagonlized by a biunitary transformation,
such as follows
\begin{equation}
\text{diag}\left(m_{e},m_{\mu},m_{\tau}\right)=\hat{M}_{\ell}=V_{L}^{\ell\dagger}M_{\ell}V_{R}.\label{eq:Ml_biunitary}
\end{equation}

In contrast, from \eqref{eq:LYukawa}, after the SSB, the neutrino
mass matrix in the basis $n_{L}=\left(\nu_{L},\nu_{R}^{c},S^{c}\right)$
is given by
\begin{equation}
\mathcal{M}_{\nu}=\begin{pmatrix}0 & B^{\top}\\
B & C
\end{pmatrix},\label{eq:Mnu_ISS}
\end{equation}
here, 
\begin{align*}
A=0;\qquad B & =\begin{pmatrix}m_{D}\\
m_{D}^{\prime}
\end{pmatrix},\qquad C=\begin{pmatrix}0 & M_{D}^{\top}\\
M_{D} & \mu
\end{pmatrix}
\end{align*}
and

\begin{equation}
m_{D}=\frac{1}{\sqrt{2}}\left(k_{1}Y+k_{2}\tilde{Y}\right),\qquad m_{D}^{\prime}=\frac{1}{\sqrt{2}}v_{L}Y_{L},\qquad M_{D}=\frac{1}{\sqrt{2}}v_{R}Y_{R}.\label{eq:Dirac_Majoran_matrices_ISS}
\end{equation}

The light neutrino mass matrix is approximated in the limit $v_{L}=0$
and $k_{2}=0$, as follows 
\begin{equation}
m_{\nu}\approx{\displaystyle m_{D}^{\top}M_{D}^{-1}\mu\left(M_{D}^{\top}\right)^{-1}m_{D}},\label{eq:light_nu_approx}
\end{equation}
where the Schur complement is used and assuming $|C|\gg|B|$.

On one hand, the neutrino mixing matrix could be approximated as follows
\cite{Catano:2012kw,PhysRevD.86.035007}
\begin{equation}
\mathcal{U}\approx\begin{pmatrix}U_{\nu} & -\frac{i}{\sqrt{2}}m_{D}^{T}M_{D}^{-1} & \frac{1}{\sqrt{2}}m_{D}^{T}M_{D}^{-1}\\
M_{D}^{-1}\mu M_{D}^{-1}m_{D}U_{\nu} & \frac{i}{\sqrt{2}}\mathbb{I} & \frac{1}{\sqrt{2}}\mathbb{I}\\
-M_{D}^{-1}m_{D}U_{\nu} & -\frac{i}{\sqrt{2}}\mathbb{I} & \frac{1}{\sqrt{2}}\mathbb{I}
\end{pmatrix}\label{eq:U_total}
\end{equation}
where $U_{\nu}$ is a unitary matrix which diagonalize the light neutrino
matrix \eqref{eq:light_nu_approx}, also we assume $M_{D}^{-1}\mu M_{D}^{-1}m_{D}\approx0$.
On the other hand, $m_{\nu}$ can be rewritten as follows
\[
m_{\nu}\approx m_{D}^{\top}\mathcal{M}^{-1}m_{D};\qquad\mathcal{M}=M_{D}\mu^{-1}M_{D}^{\top},
\]

and as a consequence of the Casas-Ibarra parametrization \cite{Casas:2001sr,PhysRevD.91.015001}
\begin{equation}
\begin{aligned}m_{D}= & V^{\dagger}\text{diag}\left(\sqrt{\mathcal{M}_{1}},\sqrt{\mathcal{M}_{2}},\sqrt{\mathcal{M}_{3}}\right)\\
 & \times R\text{diag}\left(\sqrt{m_{\nu_{1}}},\sqrt{m_{\nu_{2}}},\sqrt{m_{\nu_{3}}}\right)U_{\nu}^{\dagger}
\end{aligned}
\label{eq:Casas-Ibarra1}
\end{equation}
where $V$ is a unitary matrix which diagonalize $\mathcal{M}$ and
R is a complex orthogonal matrix. In the simple case of $M_{D}=\text{diag}\left(M_{D1},M_{D2},M_{D3}\right)$,
$\mu=\mu_{X}\mathbb{I}$ and $R=\mathbb{I}$, we have
\begin{equation}
\mathcal{M}=\frac{1}{\mu_{X}}\text{diag}\left(M_{D1}^{2},M_{D2}^{2},M_{D3}^{2}\right),\qquad V=\mathbb{I}.\label{eq:M_ISS_CI}
\end{equation}

and 
\begin{equation}
\begin{aligned}m_{D}= & \frac{1}{\sqrt{\mu_{X}}}\begin{pmatrix}\sqrt{m_{\nu_{1}}}M_{D1} & 0 & 0\\
0 & \sqrt{m_{\nu_{2}}}M_{D2} & 0\\
0 & 0 & \sqrt{m_{\nu_{3}}}M_{D3}
\end{pmatrix}U_{\nu}^{\dagger}.\end{aligned}
\label{eq:Casas-Ibarra2}
\end{equation}
As a consequence, 
\begin{align}
M_{D}^{-1}m_{D} & =\frac{1}{\sqrt{\mu_{X}}}\begin{pmatrix}\sqrt{m_{\nu_{1}}} & 0 & 0\\
0 & \sqrt{m_{\nu_{2}}} & 0\\
0 & 0 & \sqrt{m_{\nu_{3}}}
\end{pmatrix}U_{\nu}^{\dagger}\label{eq:Minv_mD}\\
m_{D}^{T}M_{D}^{-1} & =\frac{1}{\sqrt{\mu_{X}}}U_{\nu}^{*}\begin{pmatrix}\sqrt{m_{\nu_{1}}} & 0 & 0\\
0 & \sqrt{m_{\nu_{2}}} & 0\\
0 & 0 & \sqrt{m_{\nu_{3}}}
\end{pmatrix}\label{eq:MDT_Minv}
\end{align}
The heavy neutrino masses are given by
\begin{align}
M_{i}^{-} & \approx M_{Di},\nonumber \\
M_{i}^{+} & \approx M_{Di}.\label{eq:Mipm}
\end{align}

\subsection{Neutrino mass basis and mixing}

The weak neutrino states are rotated into the physical states $n^{\prime}=\left(\nu,N_{i}^{-},N_{i}^{+}\right)$
as follows
\begin{align*}
n_{L}^{\prime}= & \mathcal{U}n_{L},\\
n_{R}^{\prime}= & \mathcal{U}^{*}n_{R},
\end{align*}
If we rewrite $\mathcal{U}$ from \eqref{eq:U_total} in terms of
block matrices as follows
\begin{equation}
\mathcal{U}=\begin{pmatrix}U_{L}\\
U_{R}\\
U_{S}
\end{pmatrix},\label{eq:U_blocks}
\end{equation}
with $U_{L}$, $U_{R}$ and $U_{S}$ $3\times9$ matrices whose definitions
could be derived from \eqref{eq:U_total}, accordingly, the diagonal
neutrino full mass matrix $\hat{\mathcal{M}}=\text{diag}\left(m_{i},M_{i}^{-},M_{i}^{+}\right)$
is given by
\begin{align}
\hat{\mathcal{\mathcal{M}}}= & \mathcal{U}^{\top}\mathcal{M}\mathcal{U}\nonumber \\
= & U_{L}^{\top}m_{D}^{\top}U_{R}+U_{R}^{\top}m_{D}U_{L}+U_{R}^{\top}M_{D}^{\top}U_{S}+U_{S}^{\top}M_{D}U_{R}+U_{S}^{\top}\mu U_{S}.\label{eq:Mnu_ISS_diag}
\end{align}
In addition, the unitary property of $\mathcal{U}$ implies the following
unitary conditions 
\begin{align*}
U_{X}U_{Y}^{\dagger} & =\begin{cases}
\mathbb{I} & X=Y\\
0 & X\neq Y
\end{cases};\qquad X,Y=L,R,S.
\end{align*}

then, from diagonal neutrino mass matrix $\hat{\mathcal{\mathcal{M}}}$
\eqref{eq:Mnu_ISS_diag} we obtain the following identities

\begin{align}
m_{D}= & U_{R}^{*}\hat{\mathcal{M}}U_{L}^{\dagger},\nonumber \\
M_{D}= & U_{S}^{*}\hat{\mathcal{M}}U_{R}^{\dagger},\nonumber \\
\mu= & U_{S}^{*}\hat{\mathcal{M}}U_{S}^{\dagger}.\label{eq:Mdiag_mDMDmu}
\end{align}

\subsection{Yukawa interactions}

The Yukawa Lagrangian \eqref{eq:LYukawa} is rewritten as follows
\[
-\mathcal{L}_{\text{Yuk}}=\mathcal{L}_{Y}^{0}+\mathcal{L_{Y}^{\pm}},
\]
where
\begin{align*}
-\mathcal{L}_{Y}^{0}= & \overline{\ell_{R}^{\prime}}\left(\phi_{1}^{0}\tilde{Y}+\phi_{2}^{0*}Y\right)\ell_{L}^{\prime}+\overline{\nu_{R}^{\prime}}\left(\ensuremath{{\displaystyle \phi_{2}^{0}\tilde{Y}+\phi_{1}^{0*}Y}}\right)\nu_{L}^{\prime}\\
 & +\overline{S}\left(\chi_{L}^{0}Y_{L}\right)\nu_{L}^{\prime}+\overline{S^{c}}\left(\chi_{R}^{0}Y_{R}\right)\nu_{R}^{\prime}+\frac{1}{2}\overline{S^{c}}\mu S+\text{h.c.}
\end{align*}
\begin{align*}
-\mathcal{L}_{Y}^{\pm}= & \overline{\ell_{R}^{\prime}}\left(\phi_{1}^{-}Y-\phi_{2}^{-}\tilde{Y}\right)\nu_{L}^{\prime}+\overline{\nu_{R}^{\prime}}\left(\phi_{2}^{+}Y-\phi_{1}^{+}\tilde{Y}\right)\ell_{L}^{\prime}\\
 & -\overline{S}\left(\chi_{L}^{+}Y_{L}\right)\ell_{L}^{\prime}-\overline{S^{c}}\left(\chi_{R}^{+}Y_{R}\right)\ell_{R}^{\prime}+\text{h.c.}
\end{align*}
Finally, the interaction lagrangian of $h_{SM}=H_{1}^{0}$ with charged
leptons and neutrinos are given by
\begin{align}
-\mathcal{L}_{h\ell\ell} & =\frac{\sqrt{2}}{k_{1}}m_{\ell}h^{SM}\overline{\ell}\ell,\label{eq:Lhll}
\end{align}
\begin{equation}
-\mathcal{L}_{hnn}=\frac{1}{\sqrt{2}k_{1}}h^{SM}\overline{n}\left[\left(\Gamma+\Gamma^{\top}\right)P_{L}+\left(\Gamma^{\dagger}+\Gamma^{*}\right)P_{R}\right]n\label{eq:Lhnn}
\end{equation}
In the limit of $\epsilon\to0$, 
\begin{align}
\Gamma\approx & U_{R}^{\top}m_{D}U_{L}=\hat{\mathcal{M}}U_{L}^{\dagger}U_{L},\label{eq:Gamma_approx}
\end{align}
and the identity $\overline{n_{j}}P_{R}n_{i}=\overline{n_{i}}P_{L}n_{j}$
is used. The . In the case of charged scalars, we have the following
interaction Lagrangian,
\begin{align}
-\mathcal{L}_{Y}^{\pm}= & \frac{\sqrt{2}}{k_{1}}G_{L}^{-}\overline{\ell}\left(T_{RL}^{\dagger}P_{R}-m_{\ell}Q_{L}P_{L}\right)n+\frac{\sqrt{2}}{v_{R}}G_{R}^{-}\overline{\ell}\left(m_{\ell}Q_{R}P_{R}-JP_{L}\right)n\nonumber \\
 & +\frac{\sqrt{2}}{k_{1}}H_{R}^{-}\overline{\ell}\left(KP_{L}-m_{\ell}Q_{R}P_{R}\right)n+\text{h.c.}\label{eq:LY_charged}
\end{align}
with the following definitions
\begin{align}
J= & T_{SR}^{\dagger}+K,\nonumber \\
Q_{L}= & V_{L}^{\ell\dagger}U_{L}\nonumber \\
Q_{R}= & V_{R}^{\ell\dagger}U_{R}^{*}\label{eq:QL_QR}
\end{align}
and
\begin{align}
K & =V_{R}^{\ell\dagger}m_{D}U_{L}=\frac{k_{1}}{\sqrt{2}}V_{R}^{\ell\dagger}YU_{L},\nonumber \\
T_{RL} & =U_{R}^{\top}m_{D}V_{L}^{\ell}=\frac{k_{1}}{\sqrt{2}}U_{R}^{\top}YV_{L}^{\ell},\nonumber \\
T_{SR} & =U_{S}^{\dagger}M_{D}V_{R}^{\ell}=\frac{v_{R}}{\sqrt{2}}U_{S}^{\dagger}Y_{R}V_{R}^{\ell}.\label{eq:ST_definitions}
\end{align}
where we consider the definitions of $m_{D}$ and $M_{D}$ from \eqref{eq:Dirac_Majoran_matrices_ISS}.
An additional set of Feynman rules associated to LFV Higgs decays
is given in the appendix \eqref{sec:Gauge_diagonalization}.

\section{Lepton flavor violation $\ell\to\ell^{\prime}\gamma$ process}

A well know result is the amplitude for the process $\ell\to\ell^{\prime}\gamma$
can be written as follows
\[
\mathcal{A}\left(\ell\to\ell^{\prime}\gamma\right)=i\overline{u_{\ell^{\prime}}}\left(p-q\right)\epsilon_{\nu}^{*}\sigma^{\nu\mu}q_{\mu}\left[B_{L}P_{L}+B_{R}P_{R}\right]u_{\ell}\left(p\right)
\]
where $p$ and $q$ are the $\ell$ and photon momentum, respectively.
Then, the width decay is given by
\[
\Gamma\left(\ell\to\ell^{\prime}\gamma\right)=\frac{m_{\ell}^{3}}{16\pi^{2}}\left(\left|B_{L}\right|^{2}+\left|B_{R}\right|^{2}\right).
\]
The Branching Ratio can be obtained by means of 
\[
\mathcal{BR}\left(\ell\to\ell^{\prime}\gamma\right)=\frac{\Gamma\left(\ell\to\ell^{\prime}\gamma\right)}{\Gamma\left(\ell\to\ell^{\prime}\nu_{\ell}\overline{\nu}_{\ell^{\prime}}\right)+\Gamma\left(\ell\to\ell^{\prime}\gamma\right)},
\]

In the DLRSM the radiative process $\ell\to\ell^{\prime}\gamma$ are
induced at one loop and new contributions arise from $W^{\prime}$and
$H_{R}^{\pm}$. For $H_{R}^{\pm}$ the form factors are given by
\begin{align}
B_{R}^{H_{R}^{\pm}} & =\frac{em_{\ell}}{16\pi^{2}m_{H_{R}^{\pm}}^{2}}\sum_{i=1}^{9}\mathcal{K}_{ai}\mathcal{K}_{bi}^{*}G\left(\frac{m_{n_{i}}^{2}}{m_{H_{R}^{\pm}}^{2}}\right),\nonumber \\
B_{L}^{H_{R}^{\pm}} & =\frac{em_{\ell^{\prime}}}{16\pi^{2}m_{H_{R}^{\pm}}^{2}}\sum_{i=1}^{9}\mathcal{K}_{ai}\mathcal{K}_{bi}^{*}G\left(\frac{m_{n_{i}}^{2}}{m_{H_{R}^{\pm}}^{2}}\right)\label{eq:BLR_meg}
\end{align}

where 
\begin{equation}
\mathcal{K}=V_{R}^{\ell\dagger}YU_{L}\label{eq:Kmathcal}
\end{equation}
and the loop function 
\begin{equation}
G\left(t\right)=\ensuremath{{\displaystyle \frac{1}{12\left(t-1\right)^{4}}}}\left(2t^{3}-6t^{2}\log\left(t\right)+3t^{2}-6t+1\right).\label{eq:meg_loop_function}
\end{equation}

The most important regime where the $H_{R}^{\pm}$ contribution is
in the limit of $t\to1$, because $M_{i}^{\pm},m_{H_{R}^{\pm}}\sim v_{R}$,
where $G\left(t\right)\approx\frac{7}{120}-\frac{t}{60}$ approaches
a constant. When, heavy neutrinos are smaller than $m_{H_{R}^{\pm}}$,
$G\left(t\right)\approx\frac{1}{12}-\frac{t}{6}$. 

For $W$ and $W^{\prime}$ bosons, neglecting $m_{\ell^{\prime}}$,
we have 
\begin{align}
B_{R}^{W} & \approx g^{2}\frac{em_{\ell}}{64\pi^{2}m_{W}^{2}}\sum_{i=1}^{3}\left(U_{\nu}\right)_{\ell^{\prime}i}\left(U_{\nu}^{*}\right)_{\ell i}F\left(\frac{m_{\nu_{i}}^{2}}{m_{W}^{2}}\right),\label{eq:BW_muegamma}\\
B_{L}^{W^{\prime}} & =g^{2}\frac{em_{\ell}}{64\pi^{2}m_{W^{\prime}}^{2}}\sum_{j=1}^{9}\left(Q_{R}\right)_{\ell^{\prime}j}\left(Q_{R}^{*}\right)_{\ell j}F\left(\frac{m_{n_{j}}^{2}}{m_{W^{\prime}}^{2}}\right),\label{eq:BWprime_muegamma}
\end{align}
where $m_{N_{j}}$ with $j=1,\dots,9$ runs over the heavy neutrino
masses $M_{i}^{-}$and $M_{i}^{+}$. The loop scalar function $F$
is defined as
\[
F\left(t\right)=\ensuremath{\ensuremath{{\displaystyle \frac{t\left(5t^{2}-6t+9\right)\log\left(t\right)}{3\left(t-1\right)^{4}}-\frac{17t^{2}-10t+17}{9\left(t-1\right)^{3}}}}}
\]
with the following limit cases
\[
F\left(t\right){}_{t\rightarrow0}\sim\frac{17}{9}+\frac{41}{9}t,\quad F\left(t\right){}_{t\rightarrow1}\sim\frac{17}{15}-\frac{3}{10}t.
\]
For the light neutrino masses, the contribution from $W$ boson becomes
negligible in the limit $x\equiv m_{\nu_{i}}^{2}/m_{W}^{2}\to0$ where
the $F\left(x\right)$ approaches a constant. Consequently, $B_{R}^{W}\approx0$
due to unitarity of $U_{\nu}$. On the other hand, the case of $W^{\prime}$
contribution is different, in this case, $x_{N}\equiv m_{N_{i}}^{2}/m_{W^{\prime}}^{2}$.
From \eqref{eq:Wprime_mass}, $m_{W^{\prime}}^{2}\propto v_{R}^{2}$,
also, we observe that $M_{i}^{-}$and $M_{i}^{+}$are of the order
of $M_{D_{i}}\propto v_{R}$ and $t\sim1$. However, in this case,
the $Q_{R}$ is form by $3\times3$ block diagonal matrices \eqref{eq:U_total}
in consequence the 
\begin{equation}
\sum_{j=1}^{9}\left(Q_{R}\right)_{\ell^{\prime}j}\left(Q_{R}^{*}\right)_{\ell j}F\left(\frac{m_{n_{j}}^{2}}{m_{W^{\prime}}^{2}}\right)=0\label{eq:QR_suppression}
\end{equation}
and the $W^{\prime}$ is suppressed. 

\section{Lepton flavor violation $H$ decays}\label{sec:LFVHD}

In general the amplitude for LFVHD is given by
\begin{equation}
\mathcal{M}\left(H_{r}\to\ell_{a}\ell_{b}\right)=-\overline{u}\left(p_{1}\right)\left(A_{L}^{r}P_{L}+A_{R}^{r}P_{R}\right)v\left(p_{2}\right),\label{eq:amplitudHrlalb}
\end{equation}

where $A_{L,R}^{r}$ are the form factors, $p_{1,2}$ are the momentum
of $\ell_{a,b}$, and $p_{r}$ the momentum of the Higgs $H_{r}^{0}$.
Also, we consider the one-shell conditions $p_{1,2}^{2}=m_{a,b}^{2}$
and $p_{r}^{2}=\left(p_{1}+p_{2}\right)^{2}=m_{r}^{2}$ the mass of
$H_{r}^{0}$. The partial width decay is given by
\begin{align}
\Gamma\left(H_{r}\to\ell_{a}\ell_{b}\right)\equiv & \Gamma\left(H_{r}^{0}\to\ell_{a}^{-}\ell_{b}^{+}\right)+\Gamma\left(H_{r}^{0}\to\ell_{a}^{^{-}}\ell_{b}^{+}\right)\nonumber \\
= & \frac{1}{8\pi m_{r}}\left[1-\left(\frac{m_{a}^{2}+m_{b}^{2}}{m_{r}^{2}}\right)\right]^{1/2}\left[1-\left(\frac{m_{a}^{2}-m_{b}^{2}}{m_{r}^{2}}\right)\right]^{1/2}\nonumber \\
\times & \left[\left(m_{r}^{2}-m_{a}^{2}-m_{b}^{2}\right)\left(\left|A_{L}^{r}\right|^{2}+\left|A_{R}^{r}\right|^{2}\right)-4m_{a}m_{b}\text{Re}\left(A_{L}^{r}A_{R}^{r*}\right)\right]\label{eq:Width_hmutau}
\end{align}

\begin{figure}
\begin{centering}
\includegraphics[scale=0.9]{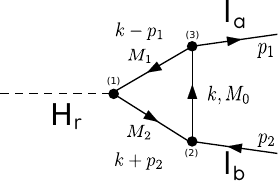}\includegraphics[scale=1.1]{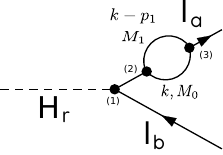}\includegraphics[scale=1.1]{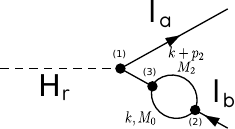}
\par\end{centering}
\caption{One loop topologies in the LFVHD, with the conventions of momentum,
label of vertexes and masses of each particle in the loop. }
\label{fig:One-loop-topologies}
\end{figure}

Considering three types of particles into the loop, charged scalars
and would-be Goldstone bosons denoted by $S^{\pm}$, charged vectors
denoted as $V^{\pm}$ and fermions denoted by $F$ , ten different
one-loop structures of Feynman diagrams appears, which are summarized
in the Table \ref{tab:Diagramas}, where the Diagram column denotes
the label to each diagram structure. These diagrams are described
with only three topologies which are vertex correction and auto energies
for each external lepton in the diagram, as it is shown in Figure
\ref{fig:One-loop-topologies}\cite{Zeleny-Mora:2021tym}.

\begin{center}
\begin{table}
\begin{centering}
\begin{tabular}{|c|c|c|c|c|c|c|c|}
\hline 
\foreignlanguage{spanish}{Diagram} & \foreignlanguage{spanish}{$P_{0}$} & \foreignlanguage{spanish}{$P_{1}$} & \foreignlanguage{spanish}{$P_{2}$} & \foreignlanguage{spanish}{Diagram} & \foreignlanguage{spanish}{$P_{0}$} & \foreignlanguage{spanish}{$P_{1}$} & \foreignlanguage{spanish}{$P_{2}$}\tabularnewline
\hline 
\hline 
\foreignlanguage{spanish}{SFF} & \foreignlanguage{spanish}{$S^{\pm}$} & \foreignlanguage{spanish}{$\overline{F_{i}}$} & \foreignlanguage{spanish}{$F_{j}$} & \foreignlanguage{spanish}{VFF} & \foreignlanguage{spanish}{$V^{\pm}$} & \foreignlanguage{spanish}{$\overline{F_{i}}$} & \foreignlanguage{spanish}{$F_{j}$}\tabularnewline
\hline 
\foreignlanguage{spanish}{FSS} & \foreignlanguage{spanish}{$F_{i}$} & \foreignlanguage{spanish}{$S^{\pm}$} & \foreignlanguage{spanish}{$S^{\mp}$} & \foreignlanguage{spanish}{FVV} & \foreignlanguage{spanish}{$F_{i}$} & \foreignlanguage{spanish}{$V^{\pm}$} & \foreignlanguage{spanish}{$V^{\mp}$}\tabularnewline
\hline 
\foreignlanguage{spanish}{FVS} & \foreignlanguage{spanish}{$F_{i}$} & \foreignlanguage{spanish}{$V^{\pm}$} & \foreignlanguage{spanish}{$S^{\mp}$} & \foreignlanguage{spanish}{FSV} & \foreignlanguage{spanish}{$F_{i}$} & \foreignlanguage{spanish}{$S^{\pm}$} & \foreignlanguage{spanish}{$V^{\mp}$}\tabularnewline
\hline 
\foreignlanguage{spanish}{FS} & \foreignlanguage{spanish}{$F_{i}$} & \foreignlanguage{spanish}{$S^{\pm}$} & \foreignlanguage{spanish}{-} & \foreignlanguage{spanish}{SF} & \foreignlanguage{spanish}{$F_{i}$} & \foreignlanguage{spanish}{-} & \foreignlanguage{spanish}{$S^{\mp}$}\tabularnewline
\hline 
\foreignlanguage{spanish}{FV} & \foreignlanguage{spanish}{$F_{i}$} & \foreignlanguage{spanish}{$V^{\pm}$} & \foreignlanguage{spanish}{-} & \foreignlanguage{spanish}{VF} & \foreignlanguage{spanish}{$F_{i}$} & \foreignlanguage{spanish}{-} & \foreignlanguage{spanish}{$V^{\mp}$}\tabularnewline
\hline 
\end{tabular}
\par\end{centering}
\caption{\foreignlanguage{spanish}{Generic diagrams which contributes to \foreignlanguage{english}{$H_{r}\to\ell_{a}\ell_{b}$},
at one loop, showing the particles inside the loop $P_{i}$ with masses
$M_{i}$ following the conventions of Figure \ref{fig:One-loop-topologies}.}}
\label{tab:Diagramas}
\end{table}
\par\end{center}

In the context of the DLRSM, the diagrams which contributes to LFVHD
are given in the Table \ref{tab:Diagrams_LFVHD}, as a consequence
the charged currents in \ref{eq:LY_charged} mediated by $S^{\pm}=G_{L}^{\pm},$$G_{R}^{\pm},H_{R}^{\pm}$
and $V^{\pm}=W^{\pm},W^{\prime^{\pm}}$and the neutrinos mixing induced
from ISS. We consider the limit of no mixing among $W-W^{\prime}$
neither $Z-Z^{\prime}$ and $k_{1}\ll v_{R}$. The total form factors
are given by 
\[
\begin{aligned}A_{L,R}^{\text{total}} & =\sum_{\Theta}A_{L,R}\left(\Theta\right)\end{aligned}
,
\]
where each contribution in Table \ref{tab:Diagrams_LFVHD}, are denote
by $\Theta$. The analytical expression for the each form factor can
be obtained following the results in\cite{Zeleny-Mora:2021tym} where
the LFVHD form factors at one loop are classified in two groups, derived
from diagrams with one neutrino in the loop or diagrams with two neutrinos
in the loop. We follow \cite{Zeleny-Mora:2021tym} to obtain the form
factors and these are shown in the Appendix \ref{sec:One-loop}.

\begin{table}
\begin{centering}
\begin{tabular}{|c|c|c|c|c|c|c|c|c|c|c|c|c|c|c|}
\hline 
No.  & $\Theta$  & $P_{0}$  & $P_{1}$  & $P_{2}$  & No.  & $\Theta$  & $P_{0}$  & $P_{1}$  & $P_{2}$  & No.  & $\Theta$  & $P_{0}$  & $P_{1}$  & $P_{2}$\tabularnewline
\hline 
\hline 
1  & SFF  & $G_{R}^{\pm}$  & $\overline{n_{i}}$  & $n_{j}$  & 11  & FSV & $n_{i}$ & $G_{R}^{\pm}$ & $W^{\prime\mp}$ & 21 & FS & $n_{i}$ & $H_{R}^{\pm}$ & ---\tabularnewline
\hline 
2  & SFF  & $G_{L}^{\pm}$  & $\overline{n_{i}}$  & $n_{j}$  & 12  & FVS & $n_{i}$ & $W^{\prime\pm}$ & $H_{R}^{\mp}$ & 22 & SF & $n_{i}$ & --- & $G_{L}^{\pm}$\tabularnewline
\hline 
3  & SFF  & $H_{R}^{\pm}$  & $\overline{n_{i}}$  & $n_{j}$  & 13  & FSV & $n_{i}$ & $H_{R}^{\pm}$ & $W^{\prime\mp}$ & 23 & SF & $n_{i}$ & --- & $G_{R}^{\pm}$\tabularnewline
\hline 
4  & VFF  & $W^{\pm}$ & $\overline{n_{i}}$  & $n_{j}$  & 14  & FSS & $n_{i}$ & $G_{R}^{\pm}$ & $G_{R}^{\mp}$ & 24 & SF & $n_{i}$ & --- & $H_{R}^{\pm}$\tabularnewline
\hline 
5  & VFF  & $W^{\prime\pm}$  & $\overline{n_{i}}$  & $n_{j}$  & 15  & FSS & $n_{i}$ & $G_{L}^{\pm}$ & $G_{L}^{\mp}$ & 25 & FV  & $n_{i}$  & $W^{\pm}$  & --- \tabularnewline
\hline 
6  & FVV  & $n_{i}$ & $W^{\pm}$ & $W^{\mp}$ & 16  & FSS & $n_{i}$ & $G_{R}^{\pm}$ & $H_{R}^{\mp}$ & 26 & FV  & $n_{i}$  & $W^{\prime\pm}$  & --- \tabularnewline
\hline 
7  & FVV  & $n_{i}$ & $W^{\prime\pm}$ & $W^{\prime\mp}$ & 17  & FSS & $n_{i}$ & $H_{R}^{\pm}$ & $G_{R}^{\mp}$ & 27 & VF  & $n_{i}$  & ---  & $W^{\pm}$ \tabularnewline
\hline 
8  & FVS & $n_{i}$ & $W^{\pm}$ & $G_{L}^{\mp}$ & 18  & FSS & $n_{i}$ & $H_{R}^{\pm}$ & $H_{R}^{\mp}$ & 28 & VF  & $n_{i}$  & ---  & $W^{\prime\pm}$ \tabularnewline
\hline 
9  & FSV & $n_{i}$ & $G_{L}^{\pm}$ & $W^{\mp}$ & 19  & FS & $n_{i}$ & $G_{L}^{\pm}$ & --- &  &  &  &  & \tabularnewline
\hline 
10  & FVS & $n_{i}$ & $W^{\prime\pm}$ & $G_{R}^{\mp}$ & 20  & FS & $n_{i}$ & $G_{R}^{\pm}$ & --- &  &  &  &  & \tabularnewline
\hline 
\end{tabular}
\par\end{centering}
\caption{Feynman diagrams with contributions to LFVHD in the DLRSM in the limit
of no gauge mixing and $k_{1}\ll v_{R}$, considering the Feynman
gauge.}
\label{tab:Diagrams_LFVHD} 
\end{table}

\section{Numerical Analysis}\label{sec:Numerical-Analysis}

The total form factors $A_{L,R}^{\text{total}}$ are function of parameters
of the potential $\alpha_{1,2,3}$, $\lambda_{1,2}$ and $\rho_{1}$,
the masses of $W^{\prime}$ and $H_{R}^{\pm}$, the scale of $v_{R}$
and the heavy neutrino masses and mixings (see Appendix \ref{sec:One-loop}).
For neutrino data the light neutrino mixings angles and mass square
differences for the Normal Ordering (NO) are given in the Table\ref{tab:Nu_data}
obtained by the NuFit collaboration\cite{Esteban:2024eli}.
\begin{table}
\begin{centering}
\begin{tabular}{|c|c|c|}
\hline 
 & \multicolumn{2}{c|}{Normal Ordering (best fit)}\tabularnewline
\hline 
\hline 
 & $\text{bfp}\pm1\sigma$ & $3\sigma$ range\tabularnewline
\hline 
$\ensuremath{\sin^{2}\theta_{12}}$ & $\ensuremath{0.308_{-0.011}^{+0.012}}$ & $\ensuremath{0.275\rightarrow0.345}$\tabularnewline
\hline 
$\ensuremath{\sin^{2}\theta_{23}}$ & $\ensuremath{0.470_{-0.013}^{+0.017}}$ & $\ensuremath{0.435\rightarrow0.585}$\tabularnewline
\hline 
$\ensuremath{\sin^{2}\theta_{13}}$ & $\ensuremath{0.02215_{-0.00058}^{+0.00056}}$ & $\ensuremath{0.02030\rightarrow0.02388}$\tabularnewline
\hline 
$\ensuremath{\delta_{\mathrm{CP}}/{}^{\circ}}$ & $\ensuremath{212_{-41}^{+26}}$ & $\ensuremath{124\rightarrow364}$\tabularnewline
\hline 
$\ensuremath{\frac{\Delta m_{21}^{2}}{10^{-5}\mathrm{eV}^{2}}}$ & $7.49_{-0.19}^{+0.19}$ & $\ensuremath{6.92\rightarrow8.05}$\tabularnewline
\hline 
$\ensuremath{\frac{\Delta m_{3\ell}^{2}}{10^{-3}\mathrm{eV}^{2}}}$ & $\ensuremath{+2.513_{-0.019}^{+0.021}}$ & $\ensuremath{+2.451\rightarrow+2.578}$\tabularnewline
\hline 
\end{tabular}
\par\end{centering}
\caption{Neutrino data for light neutrino mixing .}\label{tab:Nu_data}
\end{table}

The model presented in Sections \ref{sec:Model} and \ref{sec:ISS}
is implemented on the Mathematica package SARAH \cite{Staub:2011dp}.
Our implementation of the DLRSM model consider the Manifest Left-Right
Symmetry with $g_{L}=g_{R}$ and it is available in the Github repository
\href{https://github.com/moiseszeleny/DLRSM}{DLRSM}. SARAH package
allow to create the model files for other external software \cite{Staub:2015kfa}
such us SPheno, enabling the computation of the model’s mass spectrum
as well as other physical observables \cite{Porod:2011nf}. The workflow
of SPheno operates through input and output files (I/O) containing
numerical data that map model parameters to physical observables.
A practical approach to scan the parameters of the model with the
help of SPheno consist in modify the parameters of the model in the
input file in different benchmarks and discriminate them by the considered
bounds such us the allowed values of the scalar masses, in particular
the SM Higgs mass $m_{h_{SM}}$. However, the problem of this approach
is that we have a multivariate parameter space and a multi-objective
function to scan and it needs a lot of computational resources and
time to find satisfactory regions. Parameter scan (PS) problem has
been explored in the context of beyond the SM (BSM) analyzing adaptative
algorithms in \cite{BREIN200542}. We consider a Marcov Chain Monte
Carlo (MCMC) algorithm for PS problem implemented in the library hep-aid
\cite{Diaz:2024sxg}. This python library automate the process of
(I/O) of SPheno and other tools like Madgraph denominated as HEP stack.
This library was used in the context of (B-L) Super Symmetric model
for constraints the parameters space using bounds of the scalar mass
spectrum \cite{Diaz:2024yfu}. The automatization of (I/O) for SPheno
allows to define objective functions such as SPheno output observables.

The parameters of the potential $\alpha_{1}$, $\alpha_{3}$, $\lambda_{1}$,
$\lambda_{2}$ and $\rho_{1}$ are related with the mass of scalar
fields. The SM-like $H_{1}^{0}$ mass depends on scalar potential
parameters and $k_{1}$ \eqref{eq:Hi0_mass}. However, the masses
for $H_{i}^{0}$ with $i=2,3,4$ depends directly over $v_{R}$, \eqref{eq:Hi0_mass},
as a result, $m_{H_{1}^{0}}\ll m_{H_{i}^{0}}\approx O\left(v_{R}\right)$.
We create as a multi-objective function in hep-aid the masses of $H_{1,2,3,4}^{0}$
computed with SPheno, and use MCMC algorithm to scan the allowed parameter
space with the constraints of SM Higgs mass $m_{h_{SM}}=125.20\pm0.11$
GeV and assuming the masses of $H_{i}^{0}$ with $i=2,3,4$ greater
than the SM Higgs mass. Spheno computes the mass spectrum corresponding
to a specific benchmark point in the parameter space; however, certain
configurations may yield nonphysical results, typically manifesting
as negative mass values.  

We found a benchmark point for the values of the scalar potential
parameters as follows,
\begin{align}
\rho_{1}= & 0.6641, & \lambda_{1}= & 6.7478, & \lambda_{2}= & 3.3884,\nonumber \\
\alpha_{1}= & 3.5455, & \alpha_{2}= & 4.6905, & \alpha_{3}= & 1.5826,\label{eq:potential_parameters_benchmarck}
\end{align}
which approximate the Higgs mass $m_{H_{1}^{0}}\approx m_{h^{SM}}$and
fullfill with $m_{H_{1}^{0}}\ll m_{H_{i}^{0}}$.

For simplicity, we consider the degenerated heavy neutrino case with
\begin{equation}
M_{i}^{-}\approx M_{i}^{+}=M=\frac{Y_{R}}{\sqrt{2}}v_{R},\label{eq:Mipm_degenerate}
\end{equation}
we consider $m_{\nu_{1}}=10^{-3}$ eV in the normal order (NO) and
fix mixing angles $\theta_{12,13,23}$ to best fit point values from
Table \ref{tab:Nu_data} and $R=\mathbb{I}$. The free parameters
for $\mathcal{BR}\left(\ell\to\ell^{\prime}\gamma\right)$and $\mathcal{BR}\left(h^{SM}\to\ell_{a}\ell_{b}\right)$
are $m_{H_{R}^{\pm}}$ \eqref{eq:mHRpm}, and $M$ \eqref{eq:Mipm_degenerate}
and $\mu_{X}$ \eqref{eq:M_ISS_CI}, \eqref{eq:Casas-Ibarra2}, then,
we consider $\mu_{X},$ $Y_{R}$ and $v_{R}$ as the free parameters
in our analysis. The obtained total form factor are substituted in
\eqref{eq:amplitudHrlalb} to obtain the partial widths of LFVHD.
Here we use the library LoopTools \cite{Hahn:1998yk} to evaluate
numerically the PV functions.

\begin{figure}
\begin{centering}
\includegraphics[scale=0.58]{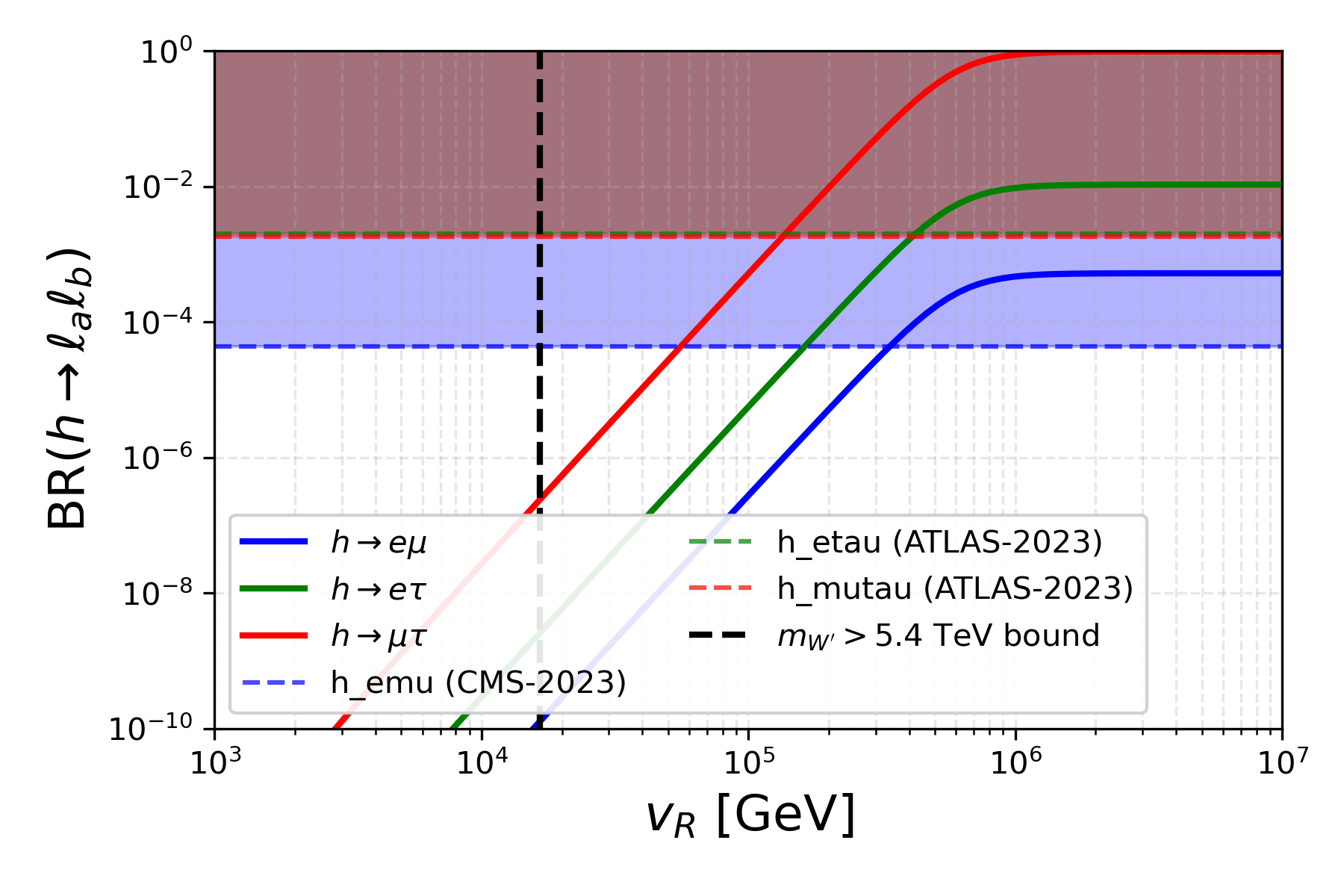}\includegraphics[scale=0.58]{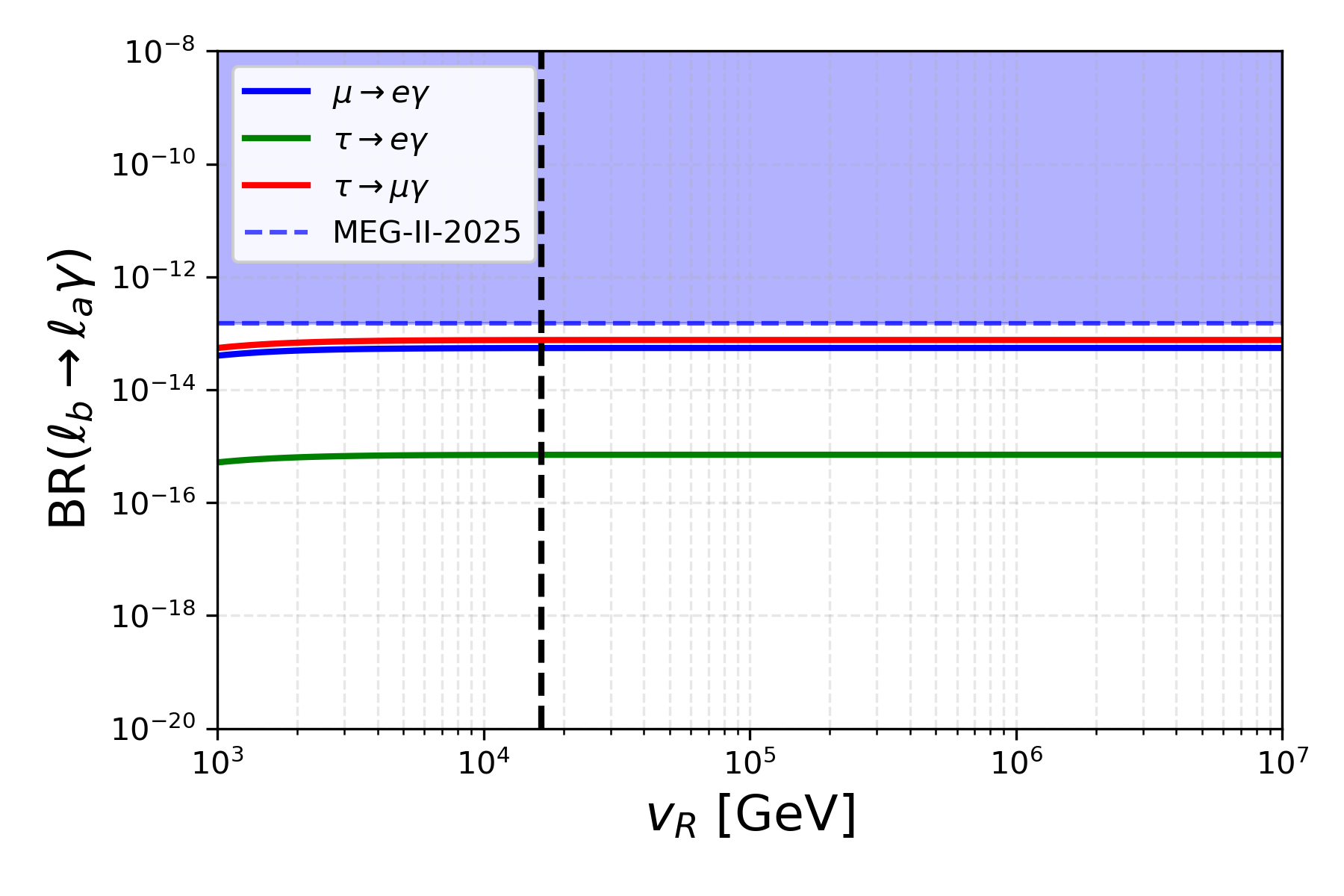}
\par\end{centering}
\caption{Behaviour of $\mathcal{BR}\left(h\to\ell_{a}\ell_{b}\right)$ left
panel and $\mathcal{BR}\left(\ell_{b}\to a\ell_{a}\right)$ right
panel. On left panel we fix $Y_{R}=0.1$, $\mu_{X}=10^{-3}$ GeV in
right panel $Y_{R}=1$ and $\mu_{X}=10^{-6}$ GeV}\label{fig:LFV_all}

\end{figure}

In Figure \ref{fig:LFV_all} we show the behavior of $\mathcal{BR}\left(h^{SM}\to\ell_{a}\ell_{b}\right)$
(left panel) and $\mathcal{BR}\left(\ell\to\ell^{\prime}\gamma\right)$
(right panel). On one hand, $\mathcal{BR}\left(\ell\to\ell^{\prime}\gamma\right)$
approaches a constant for large $v_{R}$, we observe $\mathcal{BR}\left(\tau\to\mu\gamma\right)$
presents the largest values but the stringent bound comes from MEG-II
2025 bound for $\mathcal{BR}\left(\mu\to e\gamma\right)$ (blue dashed
line) on Table \ref{tab:LFV_experimental_bounds}, $\mathcal{BR}\left(\mu\to e\gamma\right)$
is near to the this upper bound. On the other hand, $\mathcal{BR}\left(h^{SM}\to\ell_{a}\ell_{b}\right)$
increase as $v_{R}$ and approaches constant for large $v_{R}$. The
largest decay correspond to $\mathcal{BR}\left(h^{SM}\to\mu\tau\right)$.
On left panel, the dashed lines correspond to the upper bounds for
each decay in Table \ref{tab:LFV_experimental_bounds}, the most stringent
bound correspond to $\mathcal{BR}\left(h^{SM}\to\mu\tau\right)$ given
by ATLAS-2023 (red dashed line) which allows $v_{R}^{\text{max}}\approx10^{5}$
GeV in this benchmark. 

\begin{figure}
\includegraphics[scale=0.6]{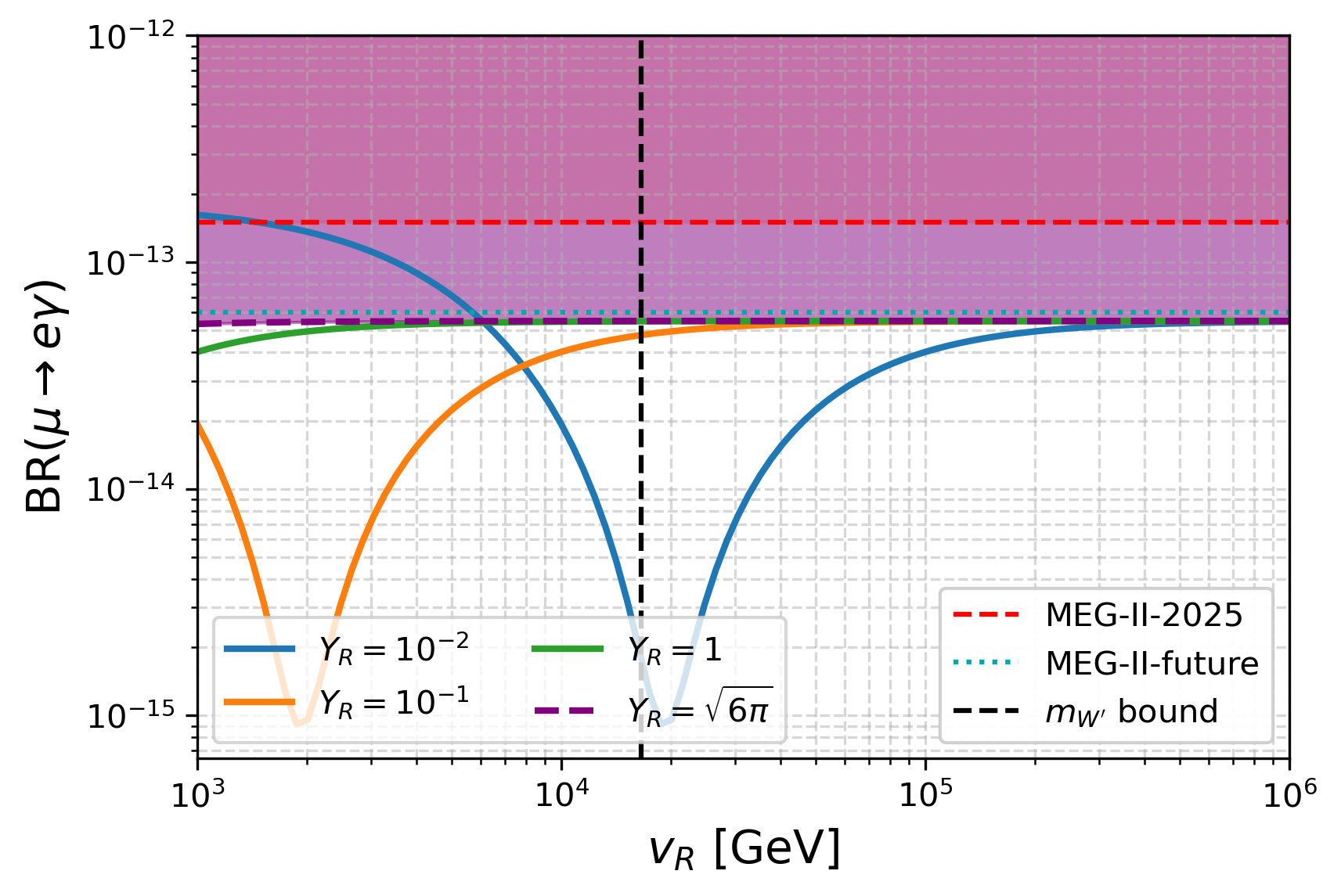}\includegraphics[scale=0.6]{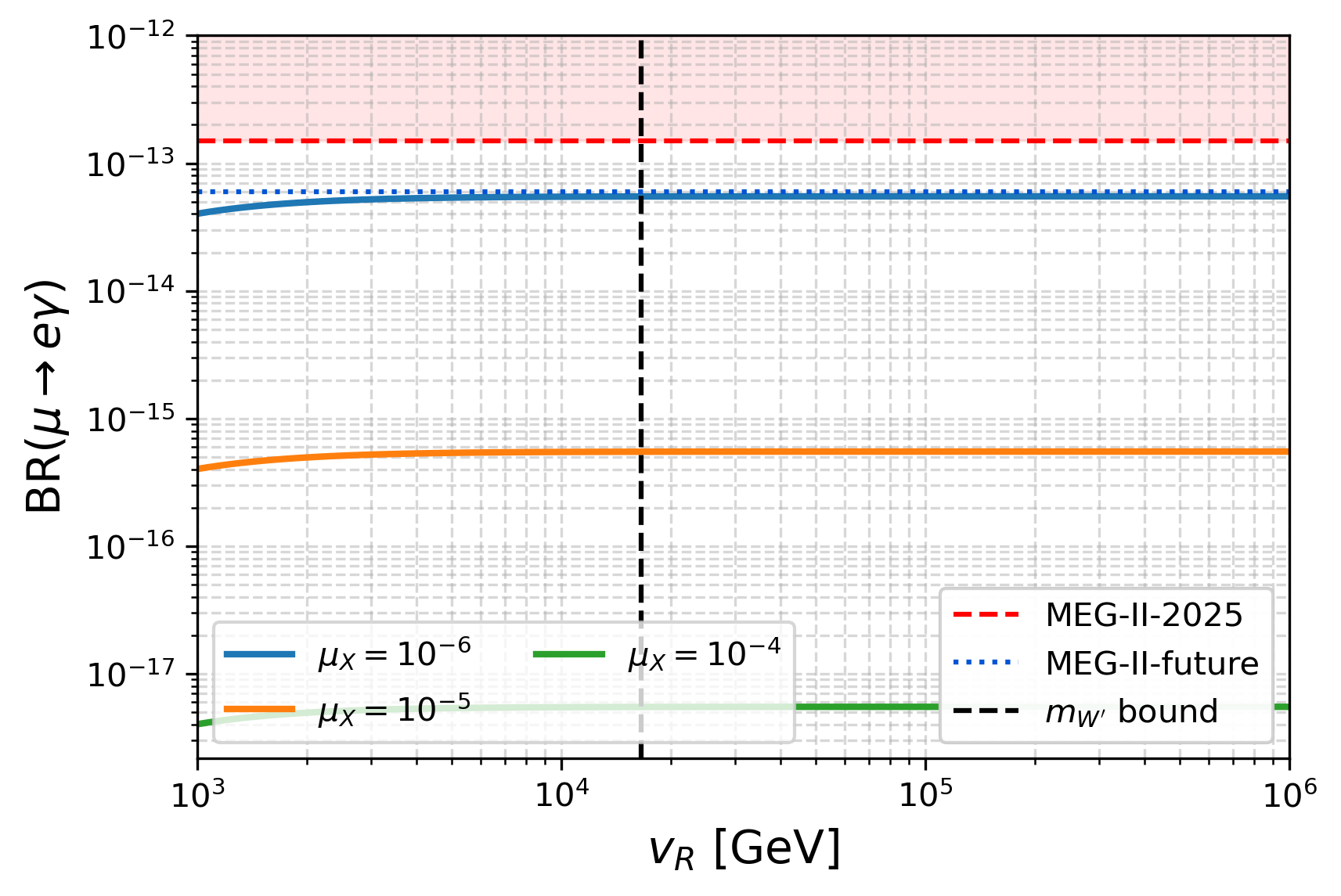}

\caption{In this figure we show the behavior of $\mathcal{BR}\left(\mu\to e\gamma\right)$as
a function of (a) $v_{R}$ and we fix $\mu_{X}=10^{-6}$ GeV, for
different values of YR and (b) $v_{R}$ and $Y_{R}=1$ and $\mu_{X}$variable.}\label{fig:meg_curves}
\end{figure}

As we observed in Figure \ref{fig:LFV_all}, the most stringent bound
correspond to $\mathcal{BR}\left(\mu\to e\gamma\right),$ then, we
explore its behavior in Figure \ref{fig:meg_curves}. The behavior
of $\mathcal{BR}\left(\mu\to e\gamma\right)$, for different values
of $Y_{R}$ (left panel) and $\mu_{X}$ (right panel) is presented.
We observe in both cases, that $\mathcal{BR}\left(\mu\to e\gamma\right)$
approaches a constant for large values of $v_{R}$, as a consequence
of the loop function $G(t)$ \eqref{eq:meg_loop_function} and the
comparable values for heavy neutrino $M_{i}^{\pm}$ and right charged
scalar $H_{R}^{\pm}$ masses. In left panel, we fix $\mu_{X}=10^{-6}$
GeV and observe in the low $v_{R}$ regime, the relatively light masses
of these mediators enhance the loop amplitudes, often pushing the
branching ratio above current experimental bounds for moderate to
large $Y_{R}$. This enhancement is particularly pronounced for $Y_{R}\sim\sqrt{6\pi}$,
where most of the parameter space becomes reachable by MEG-II projected
sensitivity. In the right panel, we consider $Y_{R}=1$ and different
values for $\mu_{X}$. In this case, $\mathcal{BR}\left(\mu\to e\gamma\right)$
exhibits a strong dependence on the lepton-number-violating parameter
$\mu_{X}$. Curve corresponding to $\mu_{X}=10^{-6}$ GeV lie significantly
near to the projected MEG-II sensitivity and the branching ratio for
curves $\mu_{X}=10^{-4},10^{-5}$ remains below experimental bounds
the complete parameter space.

\begin{figure}
\includegraphics[scale=0.6]{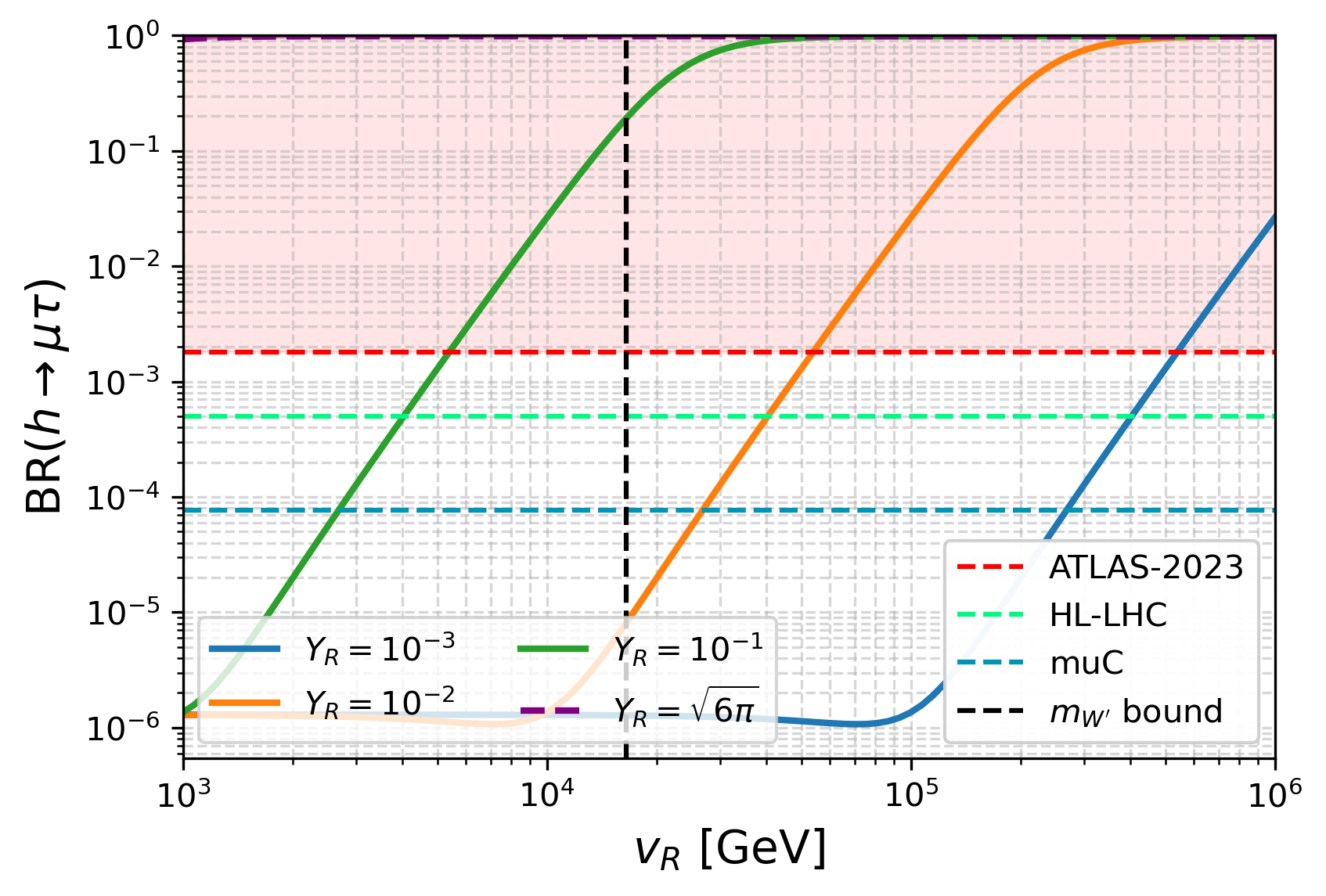}\includegraphics[scale=0.6]{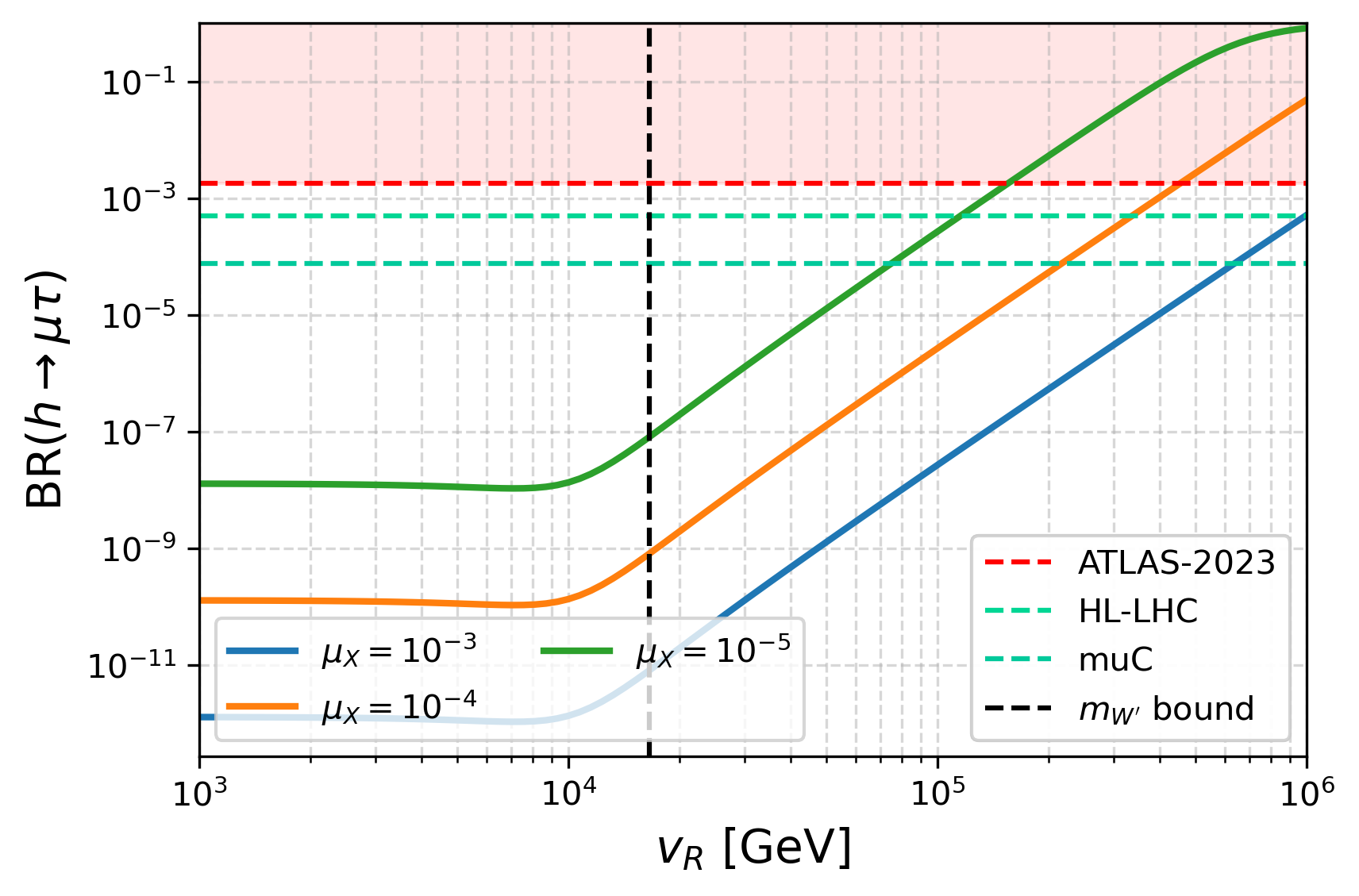}

\caption{In this figure we show the behavior of $\mathcal{BR}\left(h\to\mu\tau\right)$as
a function of (a) $v_{R}$ and we fix $\mu_{X}=10^{-6}$ GeV, for
different values of $Y_{R}$ and (b) $v_{R}$ with $Y_{R}=1$ and
$\mu_{X}$ variable.}\label{fig:hmutau_curve}

\end{figure}

On the other hand, in Figure \ref{fig:hmutau_curve}, we show $\mathcal{BR}\left(h\to\mu\tau\right)$
as a function of $v_{R}$ for different values of $Y_{R}$ (left panel)
and $\mu_{X}$ (right panel). For the left panel, we observe, the
branching ratio $\mathcal{BR}\left(h\to\mu\tau\right)$ is not sensitive
to the right-handed Yukawa coupling $Y_{R}$, particularly when the
lepton-number-violating parameter $\mu_{X}$ is fixed at a small value
such as $10^{-6}$ GeV. The plot reveals that increasing $Y_{R}$
move the branching ratio to a large right-handed neutrino mass $M$
scale ($v_{R}$). For the right panel, the branching ratio $\mathcal{BR}\left(h\to\mu\tau\right)$
is strongly influenced by the lepton-number-violating parameter $\mu_{X}$.
For fixed $Y_{R}=10^{-2}$, the plot shows that smaller values of
$\mu_{X}$ lead to a significant enhancement of the branching ratio
across the entire range of $v_{R}$. In particular, the curve corresponding
to $\mu_{X}=10^{-5}$ GeV reaches or exceeds the projected sensitivities
of HL-LHC and future muon collider for moderate $v_{R}$, while $\mu_{X}=10^{-3}$
result in suppressed branching ratios that remain below current experimental
bounds.

A recent study\cite{Lichtenstein:2025pxs} has revised the constraints
of the $Z^{\prime}$ and $W^{\prime}$masses from resonant dilepton
searches at LHC, treating the right-handed gauge coupling $g_{R}$
as a free parameter, this analysis concludes that $m_{Z^{\prime}}\gtrsim5$
TeV, $m_{W^{\prime}}\gtrsim3$ TeV, and $v_{R}$ in the range of $5$-$10$
TeV. On the experimental side, CMS collaboration have found stringent
lower bound for the $W^{\prime}$ mass given by $m_{W^{\prime}}>4.7$
TeV for heavy neutrino mass of half of the $W^{\prime}$ and $m_{W^{\prime}}>5.4$
TeV for heavy neutrinos mass around of $200$ GeV, which implies $v_{R}>1.65\times10^{4}$
GeV from \eqref{eq:Wprime_mass}, which is shown as a vertical black
line in Figures \ref{fig:LFV_all}, \ref{fig:meg_curves} and \ref{fig:hmutau_curve}.
In addition, the study of Keung-Senjanovic process in \cite{ThomasArun:2021rwf}
have found that LHC could be sensible of $W^{\prime}$mass up to $\sim6$
TeV associated to $v_{R}=1.84\times10^{4}$ GeV. 

\begin{figure}
\includegraphics[scale=0.5]{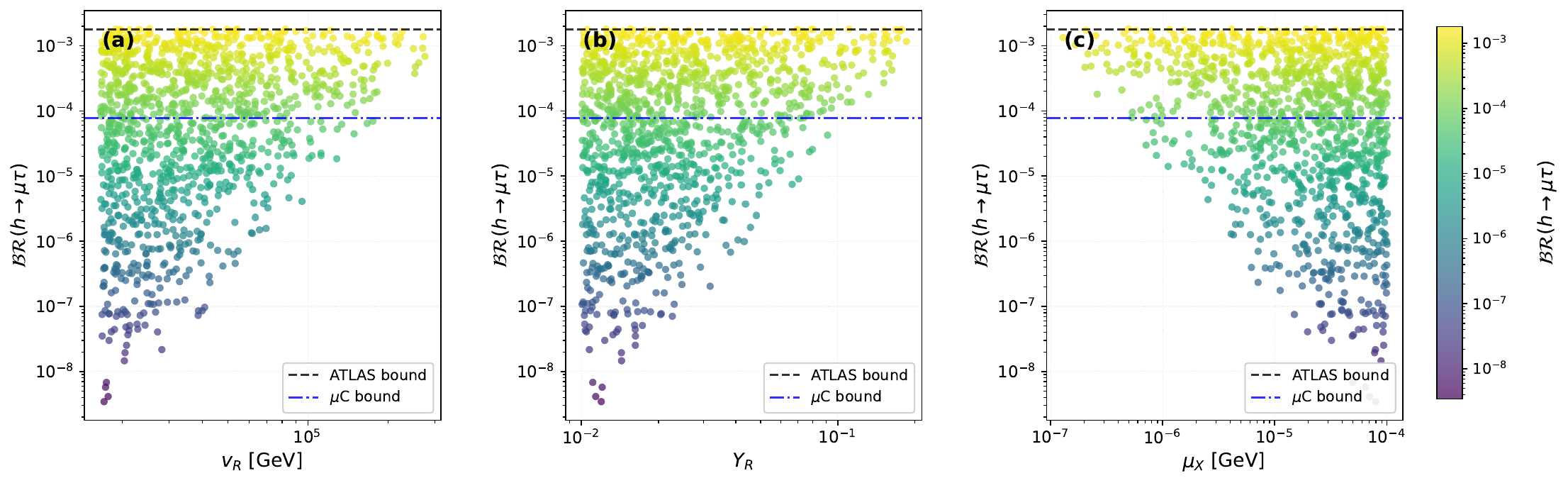}

\caption{The correlation of $\mathcal{BR}\left(h\to\mu\tau\right)$ with (a)
$v_{R}$, (b) $Y_{R}$ and (c) $\mu_{X}$ in the allowed parameter
space.}\label{fig:correlation_LFVHD}

\end{figure}

As a complement, we scan over the free parameters $Y_{R}\in[10^{-2},\sqrt{6\pi}]$,
$\mu_{X}\in[10^{-8},10^{-4}]$ GeV and $v_{R}\in[1.65\times10^{4},10^{6}]$
GeV to found the allowed parameter space by $h\to\mu\tau$, $\mu\to e\gamma$,
the perturvativity of $Y_{R}<\sqrt{6\pi}$ and the $W^{\prime}$mass
lower bound. In Figure \ref{fig:correlation_LFVHD}, we plot the allowed
parameter space and its correlation with $\mathcal{BR}\left(h\to\mu\tau\right)$.
We observe a similar correlation of $\mathcal{BR}\left(h\to\mu\tau\right)$
with $Y_{R}$ (Heavy neutrino mass $M$) in panel (a) and $v_{R}$
panel (b) and inverted correlation with $\mu_{X}$ panel (c). The
analysis yielded maximum values for $v_{R}^{\text{max}}\approx2.7\times10^{5}$
GeV and $Y_{R}^{\max}\approx2\times10^{-1}$ in consequence $M^{\max}\sim3.8\times10^{4}$
GeV.

\section{Conclusions}\label{sec:Conclusions}

In this work, we have studied LFV processes, especifically the radiative
decay $\mu\to e\gamma$ and the Higgs decay $h\to\mu\tau$ within
the framework of the DLRSM extended by an ISS mechanism. This model
provides a minimal scalar sector and avoids the stringent constraints
associated with doubly charged scalars, while still accommodating
neutrino masses and LFV signatures. 

We constructed the full neutrino mass matrix in the ISS framework
and performed its diagonalization, expressing the mixing matrices
in terms of physical parameters. The Casas--Ibarra parametrization
was used to connect the light neutrino masses and mixing to the heavy
sector, allowing for a consistent treatment of LFV observables.

The LFV rates were computed at one loop, including contributions from
charged scalars, gauge bosons and heavy neutrinos. We found that the
dominant contributions to $\mu\to e\gamma$ arise from charged scalar
loops. For $h\to\mu\tau$, multiple one-loop topologies contribute,
and we classified them systematically using vertex and self-energy
corrections. In both cases, the W\textasciiacute{} contributions are
suppressed due to mixing and mass hierarchies.

To explore the viable parameter space, we implemented the DLRSM in
SARAH and interfaced it with SPheno and hep-aid. Using a Markov Chain
Monte Carlo (MCMC) algorithm, we performed a multi-objective parameter
scan constrained by the SM Higgs mass and the requirement that additional
scalars lie above the electroweak scale. A benchmark point satisfying
these conditions was identified for scalar potential parameters. At
this benchmark point, the LFV branching ratios align with both current
and anticipated experimental sensitivities across the allowed parameter
space for $Y_{R}$, $v_{R}$ and $\mu_{X}$. As a result, an upper
bound on the right-handed scale $v_{R}$ was determined.

Our results demonstrate that the DLRSM with inverse seesaw can accommodate
observable LFV Higgs decays and radiative lepton transitions, while
remaining consistent with neutrino data and collider constraints.
Future experiments like HL-LHC and muon collider could probe these
signatures, offering a window into the flavor structure and mass generation
mechanisms beyond the Standard Model.

\begin{center}
\textbf{Acknowledgments} 
\par\end{center}

The research presented herein has been supported by the UNAM Postdoctoral
Program (POSDOC) and the PAPIIT project IN105825.

\appendix

\section{Neutral gauge boson matrix diagonalization}\label{sec:Gauge_diagonalization}

The neutral gauge boson mass matrix $M_{Z}^{2}$ in \eqref{eq:MZ_mass_matrix}
can be reduce into a block diagonal matrix by 
\begin{equation}
R=\begin{pmatrix}\sin\theta_{W} & -\cos\theta_{W} & 0\\
\sin\theta_{W} & \sin\theta_{W}\tan\theta_{W} & -\frac{\sqrt{\cos\left(2\theta_{W}\right)}}{\cos\theta_{W}}\\
\sqrt{\cos\left(2\theta_{W}\right)} & \sqrt{\cos\left(2\theta_{W}\right)}\tan\theta_{W} & \tan\theta_{W}
\end{pmatrix}\label{eq:R}
\end{equation}
with the following definitions
\begin{align}
e= & g\sin\theta_{W},\nonumber \\
\frac{1}{e^{2}}= & \frac{2}{g^{2}}+\frac{1}{g_{B-L}^{2}},\label{eq:ggBL_e_relations}
\end{align}
Then, the obtained block diagonal matrix
\begin{align*}
M_{0}^{2}= & R^{\top}M_{Z}^{2}R\\
= & \begin{pmatrix}0 & 0 & 0\\
0 & \frac{g^{2}\left(k_{1}^{2}+k_{2}^{2}\right)}{4\cos^{2}\theta_{W}} & -\frac{gg_{B-L}\left(k_{1}^{2}+k_{2}^{2}\right)}{2\cos\theta_{W}\tan\left(2\theta_{W}\right)}\\
0 & -\frac{gg_{B-L}\left(k_{1}^{2}+k_{2}^{2}\right)}{2\cos\theta_{W}\tan\left(2\theta_{W}\right)} & \frac{g^{2}v_{R}^{2}\cos^{2}\theta_{W}}{4\cos\left(2\theta_{W}\right)}+\frac{g^{2}\left(k_{1}^{2}+k_{2}^{2}\right)\cos\left(2\theta_{W}\right)}{4\cos^{2}\theta_{W}}
\end{pmatrix}
\end{align*}
could it be diagonalized by a rotation matrix $O\left(\zeta\right)$
over the angle $\zeta$ given by
\[
O\left(\zeta\right)=\begin{pmatrix}1 & 0 & 0\\
0 & \cos\left(\zeta\right) & \sin\left(\zeta\right)\\
0 & -\sin\left(\zeta\right) & \cos\left(\zeta\right)
\end{pmatrix},
\]
\begin{align}
\tan\left|2\zeta\right|\approx & \ensuremath{{\displaystyle \frac{4g_{B-L}\left(k_{1}^{2}+k_{2}^{2}\right)\cos\left(2\theta_{W}\right)}{gv_{R}^{2}\cos^{3}\theta_{W}\tan\left(2\theta_{W}\right)}}}.\label{eq:Z_mixing_angle}
\end{align}
In addition, with $R_{Z}=O\left(\zeta\right)R$ the neutral gauge
boson matrix \eqref{eq:MZ_mass_matrix}, is diagonalized by 
\begin{equation}
\hat{M}_{Z}^{2}=R_{Z}^{\top}M_{Z}^{2}R_{Z}.\label{eq:MZ_diagonal}
\end{equation}

\section{Feynman Rules}\label{sec:Feynman-Rules}

In this appendix we extract the coefficients in the Lagrangian related
with the interactions in the LFV Higgs decays, in the Feynman gauge.
We consider the limit $v_{L}=k_{2}=0$, we use the limit expressions
for the mixing among scalar fields where $k_{1}\ll v_{R}$ and assume
no mixing among charged and neutral gauge bosons, $\left(\xi=\zeta=0\right)$.
Consider the definitions
\begin{align*}
\alpha_{13}= & \alpha_{1}+\alpha_{3}\\
\alpha_{12}= & \alpha_{1}+\alpha_{2}\\
\alpha_{23}= & \alpha_{2}-\alpha_{3}\\
\lambda_{12}= & \lambda_{1}+\lambda_{2}\\
\lambda_{2356}= & \lambda_{2}-4\lambda_{3}-\lambda_{5}-\lambda_{6}
\end{align*}

then, the interactions of $H_{1}^{0}$ with $W_{1,2}^{\pm}$ is given
by
\[
\begin{array}{|c|c|}
\hline \textbf{Interaction} & \textbf{Coefficient}\\
\hline \ensuremath{W^{+}W^{-}H_{1}^{0}} & \ensuremath{\frac{g^{2}k_{1}}{2}}\\
\hline \ensuremath{W^{\prime+}W^{\prime-}H_{1}^{0}} & \frac{g^{2}k_{1}\left(2\rho_{1}-\alpha_{13}\right)}{4\rho_{1}}
\\\hline \end{array}
\]
The interactions of $H_{1}^{0}$ with $W^{\pm},W^{\prime\pm}$ and
charged Goldstones or charged scalars are given by
\[
\begin{array}{|c|c|}
\hline \textbf{Interaction} & \textbf{Coefficient}\\
\hline \ensuremath{W^{+}G_{L}^{-}H_{1}^{0}} & -\frac{g}{2}\left(p\left(G_{L}^{-}\right)-p\left(H_{1}^{0}\right)\right)\\
\hline \ensuremath{W^{\prime+}G_{R}^{-}H_{1}^{0}} & -\frac{g\left(\alpha_{13}-2\rho_{1}\right)}{4\rho_{1}}\frac{k_{1}}{v_{R}}\left(p\left(G_{R}^{-}\right)-p\left(H_{1}^{0}\right)\right)\\
\hline \ensuremath{W^{\prime+}H_{R}^{-}H_{1}^{0}} & -\frac{g}{2}\left(p\left(H_{R}^{-}\right)-p\left(H_{1}^{0}\right)\right)
\\\hline \end{array}
\]

the interaction of $H_{1}^{0}$ with charged scalars and Goldstones
is given by 
\[
\begin{array}{|c|c|}
\hline \textbf{Interaction} & \textbf{Coefficient}\\
\hline \ensuremath{G_{R}^{\pm}G_{R}^{\mp}H_{1}^{0}} & \ensuremath{-\frac{\left(-4\rho_{1}\lambda_{12}+\alpha_{13}^{2}\right)}{\rho_{1}}\frac{k_{1}^{3}}{v_{R}^{2}}}\\
\hline \ensuremath{G_{L}^{\pm}G_{L}^{\mp}H_{1}^{0}} & \ensuremath{-\frac{k_{1}\left(-4\rho_{1}\lambda_{12}+\alpha_{13}^{2}\right)}{\rho_{1}}}\\
\hline \ensuremath{G_{R}^{\pm}H_{R}^{\mp}H_{1}^{0}} & \ensuremath{v_{R}\alpha_{23}}\\
\hline \ensuremath{H_{L}^{+}H_{L}^{-}H_{1}^{0}} & \ensuremath{k_{1}\left(2\alpha_{12}-\frac{\rho_{2}}{\rho_{1}}\alpha_{13}\right)}\\
\hline \ensuremath{H_{R}^{+}H_{R}^{-}H_{1}^{0}} & k_{1}\left(2\left(\alpha_{23}+2\lambda_{12}\right)-\frac{1}{\rho_{1}}\alpha_{12}\alpha_{13}\right)
\\\hline \end{array}
\]
The interactions of $W^{\pm}$ and $W^{\prime\pm}$with leptons is
given by
\[
\begin{array}{|c|c|}
\hline \textbf{Interaction} & \textbf{Coefficient}\\
\hline \ensuremath{W^{+}\overline{n_{i}}\ell_{a}} & \frac{g}{2}\gamma^{\mu}P_{L}Q_{L,ai}^{*}\\
\hline \ensuremath{W^{-}\overline{\ell_{a}}n_{i}} & \frac{g}{2}\gamma^{\mu}P_{L}Q_{L,ai}\\
\hline \ensuremath{\ensuremath{W^{\prime+}\overline{n_{i}}\ell_{a}}} & \ensuremath{\frac{g}{2}}\gamma^{\mu}P_{R}Q_{R,ai}^{*}\\
\hline \ensuremath{W^{\prime-}\overline{\ell_{a}}n_{i}} & \ensuremath{\frac{g}{2}}\gamma^{\mu}P_{R}Q_{R,ai}
\\\hline \end{array}
\]

\section{One loop form factors of LFVZD}\label{sec:One-loop}

In this appendix we compile the results for the functions $\mathcal{H}_{\Theta}^{i}$
and $\mathcal{I}_{\Theta}^{i}$ on which the form factors $\Omega_{L,R}$
and $\Lambda_{L,R},$depend. The definition in the following depends
on the Passarino Veltman functions \cite{THAO2017159},
\[
\begin{gathered}B_{0}^{(1)}=\frac{1}{i\pi^{2}}\int d^{D}k\frac{1}{D_{0}D_{1}};\\
B_{0}^{(2)}=\frac{1}{i\pi^{2}}\int d^{D}k\frac{1}{D_{0}D_{2}};\\
B_{0}^{(12)}=\frac{1}{i\pi^{2}}\int d^{D}k\frac{1}{D_{1}D_{2}};\\
C_{0}=\frac{1}{i\pi^{2}}\int d^{D}k\frac{1}{D_{1}D_{0}D_{2}};\\
C^{\mu}=\frac{1}{i\pi^{2}}\int d^{D}k\frac{k^{\mu}}{D_{1}D_{0}D_{2}}=p_{1}^{\mu}C_{1}+p_{2}^{\mu}C_{2}
\end{gathered}
\]

where, $D_{0}=k^{2}-M_{0}^{2},\quad D_{1}=\left(k-p_{1}\right)^{2}-M_{1}^{2},\quad D_{2}=\left(k+p_{2}\right)^{2}-M_{2}^{2}.$
We consider the dimensional regularization, and $D$ is the dimension.

\subsection{One fermion in the loop}

\subsubsection{Burbujas}

For bubbles of type $n_{i}S$ or $Sn_{i}$, with $S=G_{L}^{\pm},G_{R}^{\pm},H_{R}^{\pm}$,
\begin{align*}
B_{0}^{\left(1\right)}= & B_{0}^{\left(1\right)}(m_{l_{a}},m_{n_{i}},m_{S})\\
B_{1}^{\left(1\right)}= & B_{0}^{\left(1\right)}(m_{l_{a}},m_{n_{i}},m_{S})\\
B_{0}^{\left(2\right)}= & B_{0}^{\left(2\right)}(m_{l_{b}},m_{n_{i}},m_{S})\\
B_{1}^{\left(2\right)}= & B_{1}^{\left(2\right)}(m_{l_{b}},m_{n_{i}},m_{S})
\end{align*}
and the form factors are given by 
\begin{align*}
A_{L}\left(n_{i}G_{L}^{\pm}\right)= & \frac{\sqrt{2}im_{l_{a}}m_{l_{b}}^{2}m_{n_{i}}\left(Q_{Lai}T_{RLib}+Q_{Lbi}^{*}T_{RLia}^{*}\right)B_{0}^{\left(1\right)}-\sqrt{2}im_{l_{a}}m_{l_{b}}^{2}\left(Q_{Lai}Q_{Lbi}^{*}m_{l_{a}}^{2}+T_{RLib}T_{RLia}^{*}\right)B_{1}^{\left(1\right)}}{8\pi^{2}k_{1}^{3}\left(m_{l_{a}}^{2}-m_{l_{b}}^{2}\right)}\\
A_{R}\left(n_{i}G_{L}^{\pm}\right)= & \frac{-\sqrt{2}im_{l_{a}}^{2}m_{l_{b}}\left(Q_{Lai}Q_{Lbi}^{*}m_{l_{b}}^{2}+T_{RLib}T_{RLia}^{*}\right)B_{1}^{\left(1\right)}+\sqrt{2}im_{l_{b}}m_{n_{i}}\left(Q_{Lai}T_{RLib}m_{l_{a}}^{2}+Q_{Lbi}^{*}T_{RLia}^{*}m_{l_{b}}^{2}\right)B_{0}^{\left(1\right)}}{8\pi^{2}k_{1}^{3}\left(m_{l_{a}}^{2}-m_{l_{b}}^{2}\right)}\\
A_{L}\left(G_{L}^{\pm}n_{i}\right)= & \frac{-\sqrt{2}im_{l_{a}}^{2}m_{l_{b}}m_{n_{i}}\left(Q_{Lbi}T_{RLia}+Q_{Lai}^{*}T_{RLib}^{*}\right)B_{0}^{\left(2\right)}-\sqrt{2}im_{l_{a}}m_{l_{b}}^{2}\left(Q_{Lbi}Q_{Lai}^{*}m_{l_{a}}^{2}+T_{RLia}T_{RLib}^{*}\right)B_{1}^{\left(2\right)}}{8\pi^{2}k_{1}^{3}\left(m_{l_{a}}^{2}-m_{l_{b}}^{2}\right)}\\
A_{R}\left(G_{L}^{\pm}n_{i}\right)= & \frac{-\sqrt{2}im_{l_{a}}^{2}m_{l_{b}}\left(Q_{Lbi}Q_{Lai}^{*}m_{l_{b}}^{2}+T_{RLia}T_{RLib}^{*}\right)B_{1}^{\left(2\right)}-\sqrt{2}im_{l_{a}}m_{n_{i}}\left(Q_{Lbi}T_{RLia}m_{l_{b}}^{2}+Q_{Lai}^{*}T_{RLib}^{*}m_{l_{a}}^{2}\right)B_{0}^{\left(2\right)}}{8\pi^{2}k_{1}^{3}\left(m_{l_{a}}^{2}-m_{l_{b}}^{2}\right)}\\
A_{L}\left(n_{i}G_{R}^{\pm}\right)= & \frac{-\sqrt{2}im_{l_{a}}^{2}m_{l_{b}}\left(J_{ai}J_{bi}^{*}+Q_{Rai}Q_{Rbi}^{*}m_{l_{b}}^{2}\right)B_{1}^{\left(1\right)}+\sqrt{2}im_{l_{b}}m_{n_{i}}\left(J_{ai}Q_{Rbi}^{*}m_{l_{b}}^{2}+Q_{Rai}J_{bi}^{*}m_{l_{a}}^{2}\right)B_{0}^{\left(1\right)}}{8\pi^{2}k_{1}v_{R}^{2}\left(m_{l_{a}}^{2}-m_{l_{b}}^{2}\right)}\\
A_{R}\left(n_{i}G_{R}^{\pm}\right)= & \frac{\sqrt{2}im_{l_{a}}m_{l_{b}}^{2}m_{n_{i}}\left(J_{ai}Q_{Rbi}^{*}+Q_{Rai}J_{bi}^{*}\right)B_{0}^{\left(1\right)}-\sqrt{2}im_{l_{a}}m_{l_{b}}^{2}\left(J_{ai}J_{bi}^{*}+Q_{Rai}Q_{Rbi}^{*}m_{l_{a}}^{2}\right)B_{1}^{\left(1\right)}}{8\pi^{2}k_{1}v_{R}^{2}\left(m_{l_{a}}^{2}-m_{l_{b}}^{2}\right)}\\
A_{L}\left(G_{R}^{\pm}n_{i}\right)= & \frac{-\sqrt{2}im_{l_{a}}^{2}m_{l_{b}}\left(J_{bi}J_{ai}^{*}+Q_{Rbi}Q_{Rai}^{*}m_{l_{b}}^{2}\right)B_{1}^{\left(2\right)}-\sqrt{2}im_{l_{a}}m_{n_{i}}\left(J_{bi}Q_{Rai}^{*}m_{l_{a}}^{2}+Q_{Rbi}J_{ai}^{*}m_{l_{b}}^{2}\right)B_{0}^{\left(2\right)}}{8\pi^{2}k_{1}v_{R}^{2}\left(m_{l_{a}}^{2}-m_{l_{b}}^{2}\right)}\\
A_{R}\left(G_{R}^{\pm}n_{i}\right)= & \frac{-\sqrt{2}im_{l_{a}}^{2}m_{l_{b}}m_{n_{i}}\left(J_{bi}Q_{Rai}^{*}+Q_{Rbi}J_{ai}^{*}\right)B_{0}^{\left(2\right)}-\sqrt{2}im_{l_{a}}m_{l_{b}}^{2}\left(J_{bi}J_{ai}^{*}+Q_{Rbi}Q_{Rai}^{*}m_{l_{a}}^{2}\right)B_{1}^{\left(2\right)}}{8\pi^{2}k_{1}v_{R}^{2}\left(m_{l_{a}}^{2}-m_{l_{b}}^{2}\right)}\\
A_{L}\left(n_{i}H_{R}^{\pm}\right)= & \frac{-\sqrt{2}im_{l_{a}}^{2}m_{l_{b}}\left(K_{ai}K_{bi}^{*}+Q_{Rai}Q_{Rbi}^{*}m_{l_{b}}^{2}\right)B_{1}^{\left(1\right)}+\sqrt{2}im_{l_{b}}m_{n_{i}}\left(K_{ai}Q_{Rbi}^{*}m_{l_{b}}^{2}+Q_{Rai}K_{bi}^{*}m_{l_{a}}^{2}\right)B_{0}^{\left(1\right)}}{8\pi^{2}k_{1}^{3}\left(m_{l_{a}}^{2}-m_{l_{b}}^{2}\right)}\\
A_{R}\left(n_{i}H_{R}^{\pm}\right)= & \frac{\sqrt{2}im_{l_{a}}m_{l_{b}}^{2}m_{n_{i}}\left(K_{ai}Q_{Rbi}^{*}+Q_{Rai}K_{bi}^{*}\right)B_{0}^{\left(1\right)}-\sqrt{2}im_{l_{a}}m_{l_{b}}^{2}\left(K_{ai}K_{bi}^{*}+Q_{Rai}Q_{Rbi}^{*}m_{l_{a}}^{2}\right)B_{1}^{\left(1\right)}}{8\pi^{2}k_{1}^{3}\left(m_{l_{a}}^{2}-m_{l_{b}}^{2}\right)}\\
A_{L}\left(H_{R}^{\pm}n_{i}\right)= & \frac{-\sqrt{2}im_{l_{a}}^{2}m_{l_{b}}\left(K_{bi}K_{ai}^{*}+Q_{Rbi}Q_{Rai}^{*}m_{l_{b}}^{2}\right)B_{1}^{\left(2\right)}-\sqrt{2}im_{l_{a}}m_{n_{i}}\left(K_{bi}Q_{Rai}^{*}m_{l_{a}}^{2}+Q_{Rbi}K_{ai}^{*}m_{l_{b}}^{2}\right)B_{0}^{\left(2\right)}}{8\pi^{2}k_{1}^{3}\left(m_{l_{a}}^{2}-m_{l_{b}}^{2}\right)}\\
A_{R}\left(H_{R}^{\pm}n_{i}\right)= & \frac{-\sqrt{2}im_{l_{a}}^{2}m_{l_{b}}m_{n_{i}}\left(K_{bi}Q_{Rai}^{*}+Q_{Rbi}K_{ai}^{*}\right)B_{0}^{\left(2\right)}-\sqrt{2}im_{l_{a}}m_{l_{b}}^{2}\left(K_{bi}K_{ai}^{*}+Q_{Rbi}Q_{Rai}^{*}m_{l_{a}}^{2}\right)B_{1}^{\left(2\right)}}{8\pi^{2}k_{1}^{3}\left(m_{l_{a}}^{2}-m_{l_{b}}^{2}\right)}
\end{align*}
for bubbles of type $n_{i}V$ or $Vn_{i}$with $V=W,W^{\prime},$the
PV functions are
\begin{align*}
B_{0}^{\left(1\right)}= & B_{0}^{\left(1\right)}(m_{l_{a}},m_{n_{i}},m_{V}),\\
B_{1}^{\left(1\right)}= & B_{0}^{\left(1\right)}(m_{l_{a}},m_{n_{i}},m_{V}),\\
B_{0}^{\left(2\right)}= & B_{0}^{\left(2\right)}(m_{l_{b}},m_{n_{i}},m_{V}),\\
B_{1}^{\left(2\right)}= & B_{1}^{\left(2\right)}(m_{l_{b}},m_{n_{i}},m_{v}),
\end{align*}
and the form factors are given by
\begin{align*}
A_{L}\left(n_{i}W^{\pm}\right)= & -\frac{\sqrt{2}iQ_{Lai}Q_{Lbi}^{*}g^{2}m_{l_{a}}m_{l_{b}}^{2}B_{1}^{\left(1\right)}}{32\pi^{2}k_{1}\left(m_{l_{a}}^{2}-m_{l_{b}}^{2}\right)}; & A_{R}\left(n_{i}W^{\pm}\right)= & -\frac{\sqrt{2}iQ_{Lai}Q_{Lbi}^{*}g^{2}m_{l_{a}}^{2}m_{l_{b}}B_{1}^{\left(1\right)}}{32\pi^{2}k_{1}\left(m_{l_{a}}^{2}-m_{l_{b}}^{2}\right)};\\
A_{L}\left(W^{\pm}n_{i}\right)= & -\frac{\sqrt{2}iQ_{Lbi}Q_{Lai}^{*}g^{2}m_{l_{a}}m_{l_{b}}^{2}B_{1}^{\left(2\right)}}{32\pi^{2}k_{1}\left(m_{l_{a}}^{2}-m_{l_{b}}^{2}\right)}; & A_{R}\left(W^{\pm}n_{i}\right)= & -\frac{\sqrt{2}iQ_{Lbi}Q_{Lai}^{*}g^{2}m_{l_{a}}^{2}m_{l_{b}}B_{1}^{\left(2\right)}}{32\pi^{2}k_{1}\left(m_{l_{a}}^{2}-m_{l_{b}}^{2}\right)};\\
A_{L}\left(n_{i}W^{\prime\pm}\right)= & -\frac{\sqrt{2}iQ_{Rai}Q_{Rbi}^{*}g^{2}m_{l_{a}}^{2}m_{l_{b}}B_{1}^{\left(1\right)}}{32\pi^{2}k_{1}\left(m_{l_{a}}^{2}-m_{l_{b}}^{2}\right)}; & A_{R}\left(n_{i}W^{\prime\pm}\right)= & -\frac{\sqrt{2}iQ_{Rai}Q_{Rbi}^{*}g^{2}m_{l_{a}}m_{l_{b}}^{2}B_{1}^{\left(1\right)}}{32\pi^{2}k_{1}\left(m_{l_{a}}^{2}-m_{l_{b}}^{2}\right)};\\
A_{L}\left(W^{\prime\pm}n_{i}\right)= & -\frac{\sqrt{2}iQ_{Rbi}Q_{Rai}^{*}g^{2}m_{l_{a}}^{2}m_{l_{b}}B_{1}^{\left(2\right)}}{32\pi^{2}k_{1}\left(m_{l_{a}}^{2}-m_{l_{b}}^{2}\right)}; & A_{R}\left(W^{\prime\pm}n_{i}\right)= & -\frac{\sqrt{2}iQ_{Rbi}Q_{Rai}^{*}g^{2}m_{l_{a}}m_{l_{b}}^{2}B_{1}^{\left(2\right)}}{32\pi^{2}k_{1}\left(m_{l_{a}}^{2}-m_{l_{b}}^{2}\right)}.
\end{align*}

\subsubsection{Triangles}

For triangles $n_{i}XY$ with $X,Y=S,V$ and $S=G_{L}^{\pm},G_{R}^{\pm},H_{R}^{\pm}$,
$V=W,W^{\prime}$, the PV functions are given by

\begin{align*}
C_{0,1,2}= & C_{0}(m_{H_{1}^{0}},m_{l_{a}},m_{l_{b}},m_{n_{i}},m_{X},m_{Y})\\
B_{0}^{\left(12\right)}= & B_{0}^{\left(12\right)}(m_{H_{1}^{0}},m_{X},m_{Y})
\end{align*}
\begin{align*}
A_{L}\left(n_{i}G_{L}^{+}G_{L}^{-}\right)= & \frac{im_{l_{a}}\left(\alpha_{13}^{2}-4\lambda_{12}\rho_{1}\right)\left(-Q_{Lai}T_{RLib}m_{n_{i}}C_{0}-Q_{Lai}Q_{Lbi}^{*}m_{l_{b}}^{2}C_{2}+T_{RLib}T_{RLia}^{*}C_{1}\right)}{8\pi^{2}\rho_{1}k_{1}}\\
A_{R}\left(n_{i}G_{L}^{+}G_{L}^{-}\right)= & \frac{im_{l_{b}}\left(\alpha_{13}^{2}-4\lambda_{12}\rho_{1}\right)\left(Q_{Lai}Q_{Lbi}^{*}m_{l_{a}}^{2}C_{1}-T_{RLib}T_{RLia}^{*}C_{2}-Q_{Lbi}^{*}T_{RLia}^{*}m_{n_{i}}C_{0}\right)}{8\pi^{2}\rho_{1}k_{1}}\\
A_{L}\left(n_{i}G_{R}^{+}G_{R}^{-}\right)= & \frac{im_{l_{b}}\left(\alpha_{13}^{2}-4\lambda_{12}\rho_{1}\right)\left(-J_{ai}J_{bi}^{*}k_{1}^{3}C_{2}-J_{ai}Q_{Rbi}^{*}k_{1}^{3}m_{n_{i}}C_{0}+Q_{Rai}Q_{Rbi}^{*}k_{1}^{3}m_{l_{a}}^{2}C_{1}\right)}{8\pi^{2}\rho_{1}v_{R}^{4}}\\
A_{R}\left(n_{i}G_{R}^{+}G_{R}^{-}\right)= & \frac{im_{l_{a}}\left(J_{ai}J_{bi}^{*}k_{1}^{3}C_{1}-Q_{Rai}J_{bi}^{*}k_{1}^{3}m_{n_{i}}C_{0}-Q_{Rai}Q_{Rbi}^{*}k_{1}^{3}m_{l_{b}}^{2}C_{2}\right)}{8\pi^{2}\rho_{1}v_{R}^{4}}\\
A_{L}\left(n_{i}H_{R}^{+}H_{R}^{-}\right)= & \frac{im_{l_{b}}\left(\alpha_{12}\alpha_{13}v_{R}^{2}+\alpha_{13}\alpha_{23}k_{1}^{2}-2\alpha_{23}\rho_{1}k_{1}^{2}-2\alpha_{23}\rho_{1}v_{R}^{2}-4\lambda_{12}\rho_{1}v_{R}^{2}\right)}{8\pi^{2}\rho_{1}k_{1}v_{R}^{2}}\\
 & \times\left(-K_{ai}K_{bi}^{*}C_{2}-K_{ai}Q_{Rbi}^{*}m_{n_{i}}C_{0}+Q_{Rai}Q_{Rbi}^{*}m_{l_{a}}^{2}C_{1}\right)\\
A_{R}\left(n_{i}H_{R}^{+}H_{R}^{-}\right)= & \frac{im_{l_{a}}\left(\alpha_{12}\alpha_{13}v_{R}^{2}+\alpha_{13}\alpha_{23}k_{1}^{2}-2\alpha_{23}\rho_{1}k_{1}^{2}-2\alpha_{23}\rho_{1}v_{R}^{2}-4\lambda_{12}\rho_{1}v_{R}^{2}\right)}{8\pi^{2}\rho_{1}k_{1}v_{R}^{2}}\\
 & \times\left(K_{ai}K_{bi}^{*}C_{1}-Q_{Rai}K_{bi}^{*}m_{n_{i}}C_{0}-Q_{Rai}Q_{Rbi}^{*}m_{l_{b}}^{2}C_{2}\right)
\end{align*}
\begin{align*}
A_{L}\left(n_{i}W^{+}W^{-}\right)= & \frac{iQ_{Lai}Q_{Lbi}^{*}g^{4}k_{1}m_{l_{a}}C_{1}}{64\pi^{2}}\\
A_{R}\left(n_{i}W^{+}W^{-}\right)= & -\frac{iQ_{Lai}Q_{Lbi}^{*}g^{4}k_{1}m_{l_{b}}C_{2}}{64\pi^{2}}\\
A_{L}\left(n_{i}W_{2}^{+}W_{2}^{-}\right)= & \frac{iQ_{Rai}Q_{Rbi}^{*}g^{4}k_{1}m_{l_{b}}\left(\alpha_{13}-2\rho_{1}\right)C_{2}}{128\pi^{2}\rho_{1}}\\
A_{R}\left(n_{i}W_{2}^{+}W_{2}^{-}\right)= & \frac{iQ_{Rai}Q_{Rbi}^{*}g^{4}k_{1}m_{l_{a}}\left(-\alpha_{13}+2\rho_{1}\right)C_{1}}{128\pi^{2}\rho_{1}}
\end{align*}
\begin{align*}
A_{L}\left(n_{i}G_{R}^{+}H_{R}^{-}\right)= & \frac{im_{l_{b}}\left(\alpha_{12}\alpha_{13}k_{1}^{2}v_{R}^{2}+\alpha_{13}^{2}k_{1}^{2}v_{R}^{2}+\alpha_{13}\alpha_{23}k_{1}^{4}+2\alpha_{23}\rho_{1}k_{1}^{2}v_{R}^{2}+2\alpha_{23}\rho_{1}v_{R}^{4}-8\lambda_{12}\rho_{1}k_{1}^{2}v_{R}^{2}\right)}{16\pi^{2}\rho_{1}k_{1}v_{R}^{4}}\\
 & \times\left(-K_{ai}J_{bi}^{*}C_{2}-K_{ai}Q_{Rbi}^{*}m_{n_{i}}C_{0}+Q_{Rai}Q_{Rbi}^{*}m_{l_{a}}^{2}C_{1}\right)\\
A_{R}\left(n_{i}G_{R}^{+}H_{R}^{-}\right)= & \frac{im_{l_{a}}\left(\alpha_{12}\alpha_{13}k_{1}^{2}v_{R}^{2}+\alpha_{13}^{2}k_{1}^{2}v_{R}^{2}+\alpha_{13}\alpha_{23}k_{1}^{4}+2\alpha_{23}\rho_{1}k_{1}^{2}v_{R}^{2}+2\alpha_{23}\rho_{1}v_{R}^{4}-8\lambda_{12}\rho_{1}k_{1}^{2}v_{R}^{2}\right)}{16\pi^{2}\rho_{1}k_{1}v_{R}^{4}}\\
 & \times\left(K_{ai}J_{bi}^{*}C_{1}-Q_{Rai}J_{bi}^{*}m_{n_{i}}C_{0}-Q_{Rai}Q_{Rbi}^{*}m_{l_{b}}^{2}C_{2}\right)\\
A_{L}\left(n_{i}H_{R}^{+}G_{R}^{-}\right)= & \frac{im_{l_{b}}\left(\alpha_{12}\alpha_{13}k_{1}^{2}v_{R}^{2}+\alpha_{13}^{2}k_{1}^{2}v_{R}^{2}+\alpha_{13}\alpha_{23}k_{1}^{4}+2\alpha_{23}\rho_{1}k_{1}^{2}v_{R}^{2}+2\alpha_{23}\rho_{1}v_{R}^{4}-8\lambda_{12}\rho_{1}k_{1}^{2}v_{R}^{2}\right)}{16\pi^{2}\rho_{1}k_{1}v_{R}^{4}}\\
 & \times\left(-J_{ai}K_{bi}^{*}C_{2}-J_{ai}Q_{Rbi}^{*}m_{n_{i}}C_{0}+Q_{Rai}Q_{Rbi}^{*}m_{l_{a}}^{2}C_{1}\right)\\
A_{R}\left(n_{i}H_{R}^{+}G_{R}^{-}\right)= & \frac{im_{l_{a}}\left(\alpha_{12}\alpha_{13}k_{1}^{2}v_{R}^{2}+\alpha_{13}^{2}k_{1}^{2}v_{R}^{2}+\alpha_{13}\alpha_{23}k_{1}^{4}+2\alpha_{23}\rho_{1}k_{1}^{2}v_{R}^{2}+2\alpha_{23}\rho_{1}v_{R}^{4}-8\lambda_{12}\rho_{1}k_{1}^{2}v_{R}^{2}\right)}{16\pi^{2}\rho_{1}k_{1}v_{R}^{4}}\\
 & \times\left(J_{ai}K_{bi}^{*}C_{1}-Q_{Rai}K_{bi}^{*}m_{n_{i}}C_{0}-Q_{Rai}Q_{Rbi}^{*}m_{l_{b}}^{2}C_{2}\right)
\end{align*}
\begin{align*}
A_{L}\left(n_{i}W^{+}G_{L}^{-}\right)= & \frac{\sqrt{2}ig^{2}m_{l_{a}}}{64\pi^{2}k_{1}}\left[Q_{Lai}Q_{Lbi}^{*}\left(2\left(m_{H_{1}^{0}}\right)^{2}-2m_{l_{a}}^{2}-m_{l_{b}}^{2}\right)C_{2}-Q_{Lai}Q_{Lbi}^{*}B_{0}^{\left(12\right)}\right.\\
 & \left.-Q_{Lbi}^{*}m_{n_{i}}\left(Q_{Lai}m_{n_{i}}+2T_{RLia}^{*}\right)C_{0}+Q_{Lbi}^{*}\left(2Q_{Lai}m_{l_{a}}^{2}+T_{RLia}^{*}m_{n_{i}}\right)C_{1}\right]\\
A_{R}\left(n_{i}W^{+}G_{L}^{-}\right)= & \frac{\sqrt{2}ig^{2}m_{l_{b}}}{64\pi^{2}k_{1}}\left(-Q_{Lai}Q_{Lbi}^{*}m_{l_{a}}^{2}C_{1}+Q_{Lbi}^{*}T_{RLia}^{*}m_{n_{i}}C_{0}+Q_{Lbi}^{*}\left(2Q_{Lai}m_{l_{a}}^{2}-T_{RLia}^{*}m_{n_{i}}\right)C_{2}\right)\\
A_{L}\left(n_{i}G_{L}^{+}W^{-}\right)= & \frac{i\sqrt{2}g^{2}m_{l_{a}}}{64\pi^{2}k_{1}}\left(-Q_{Lai}T_{RLib}m_{n_{i}}C_{0}-Q_{Lai}Q_{Lbi}^{*}m_{l_{b}}^{2}C_{2}-Q_{Lai}\left(T_{RLib}m_{n_{i}}-2Q_{Lbi}^{*}m_{l_{b}}^{2}\right)C_{1}\right)\\
A_{R}\left(n_{i}G_{L}^{+}W^{-}\right)= & \frac{i\sqrt{2}g^{2}m_{l_{b}}}{64\pi^{2}k_{1}}\left[Q_{Lai}Q_{Lbi}^{*}\left(2\left(m_{H_{1}^{0}}\right)^{2}-m_{l_{a}}^{2}-2m_{l_{b}}^{2}\right)C_{1}+Q_{Lai}Q_{Lbi}^{*}B_{0}^{\left(12\right)}\right.\\
 & \left.+Q_{Lai}m_{n_{i}}\left(2T_{RLib}+Q_{Lbi}^{*}m_{n_{i}}\right)C_{0}+Q_{Lai}\left(T_{RLib}m_{n_{i}}+2Q_{Lbi}^{*}m_{l_{b}}^{2}\right)C_{2}\right]\\
A_{L}\left(n_{i}W_{2}^{+}G_{R}^{-}\right)= & \frac{\sqrt{2}i\left(\alpha_{13}-2\rho_{1}\right)g^{2}k_{1}m_{l_{b}}}{128\pi^{2}\rho_{1}v_{R}^{2}}\left(-J_{ai}Q_{Rbi}^{*}m_{n_{i}}C_{0}+Q_{Rai}Q_{Rbi}^{*}m_{l_{a}}^{2}C_{1}+Q_{Rbi}^{*}\left(J_{ai}m_{n_{i}}-2Q_{Rai}m_{l_{a}}^{2}\right)C_{2}\right)\\
A_{R}\left(n_{i}W_{2}^{+}G_{R}^{-}\right)= & \frac{i\sqrt{2}\left(\alpha_{13}-2\rho_{1}\right)g^{2}k_{1}m_{l_{a}}}{128\pi^{2}\rho_{1}v_{R}^{2}}\left[-Q_{Rai}Q_{Rbi}^{*}\left(2\left(m_{H_{1}^{0}}\right)^{2}-2m_{l_{a}}^{2}-m_{l_{b}}^{2}\right)C_{2}+Q_{Rai}Q_{Rbi}^{*}B_{0}^{\left(12\right)}\right.\\
 & \left.+Q_{Rbi}^{*}m_{n_{i}}\left(2J_{ai}+Q_{Rai}m_{n_{i}}\right)C_{0}-Q_{Rbi}^{*}\left(J_{ai}m_{n_{i}}+2Q_{Rai}m_{l_{a}}^{2}\right)C_{1}\right]\\
A_{L}\left(n_{i}G_{R}^{+}W_{2}^{-}\right)= & -\frac{i\sqrt{2}\left(\alpha_{13}-2\rho_{1}\right)g^{2}k_{1}m_{l_{b}}}{128\pi^{2}\rho_{1}v_{R}^{2}}\left[Q_{Rai}Q_{Rbi}^{*}\left(2\left(m_{H_{1}^{0}}\right)^{2}-m_{l_{a}}^{2}-2m_{l_{b}}^{2}\right)C_{1}+Q_{Rai}Q_{Rbi}^{*}B_{0}^{\left(12\right)}\right.\\
 & \left.+Q_{Rai}m_{n_{i}}\left(2J_{bi}^{*}+Q_{Rbi}^{*}m_{n_{i}}\right)C_{0}+Q_{Rai}\left(J_{bi}^{*}m_{n_{i}}+2Q_{Rbi}^{*}m_{l_{b}}^{2}\right)C_{2}\right]\\
A_{R}\left(n_{i}G_{R}^{+}W_{2}^{-}\right)= & \frac{i\sqrt{2}\left(\alpha_{13}-2\rho_{1}\right)g^{2}k_{1}m_{l_{a}}}{128\pi^{2}\rho_{1}v_{R}^{2}}\left(Q_{Rai}J_{bi}^{*}m_{n_{i}}C_{0}+Q_{Rai}Q_{Rbi}^{*}m_{l_{b}}^{2}C_{2}+Q_{Rai}\left(J_{bi}^{*}m_{n_{i}}-2Q_{Rbi}^{*}m_{l_{b}}^{2}\right)C_{1}\right)\\
A_{L}\left(n_{i}W_{2}^{+}H_{R}^{-}\right)= & \frac{i\sqrt{2}\left(\alpha_{13}k_{1}^{2}+2\rho_{1}v_{R}^{2}\right)g^{2}m_{l_{b}}}{128\pi^{2}\rho_{1}k_{1}v_{R}^{2}}\left(K_{ai}Q_{Rbi}^{*}m_{n_{i}}C_{0}-Q_{Rai}Q_{Rbi}^{*}m_{l_{a}}^{2}C_{1}-Q_{Rbi}^{*}\left(K_{ai}m_{n_{i}}-2Q_{Rai}m_{l_{a}}^{2}\right)C_{2}\right)\\
A_{R}\left(n_{i}W_{2}^{+}H_{R}^{-}\right)= & \frac{i\sqrt{2}\left(\alpha_{13}k_{1}^{2}+2\rho_{1}v_{R}^{2}\right)g^{2}m_{l_{a}}}{128\pi^{2}\rho_{1}k_{1}v_{R}^{2}}\left[Q_{Rai}Q_{Rbi}^{*}\left(2\left(m_{H_{1}^{0}}\right)^{2}-2m_{l_{a}}^{2}-m_{l_{b}}^{2}\right)C_{2}-Q_{Rai}Q_{Rbi}^{*}B_{0}^{\left(12\right)}\right.\\
 & \left.-Q_{Rbi}^{*}m_{n_{i}}\left(2K_{ai}+Q_{Rai}m_{n_{i}}\right)C_{0}+Q_{Rbi}^{*}g^{2}m_{l_{a}}\left(K_{ai}m_{n_{i}}+2Q_{Rai}m_{l_{a}}^{2}\right)C_{1}\right]\\
A_{L}\left(n_{i}H_{R}^{+}W_{2}^{-}\right)= & \frac{i\sqrt{2}\left(\alpha_{13}k_{1}^{2}+2\rho_{1}v_{R}^{2}\right)g^{2}m_{l_{b}}}{128\pi^{2}\rho_{1}k_{1}v_{R}^{2}}\left[Q_{Rai}Q_{Rbi}^{*}\left(2\left(m_{H_{1}^{0}}\right)^{2}-m_{l_{a}}^{2}-2m_{l_{b}}^{2}\right)C_{1}+Q_{Rai}Q_{Rbi}^{*}B_{0}^{\left(12\right)}\right.\\
 & \left.+Q_{Rai}g^{2}m_{l_{b}}m_{n_{i}}\left(2K_{bi}^{*}+Q_{Rbi}^{*}m_{n_{i}}\right)C_{0}+Q_{Rai}\left(K_{bi}^{*}m_{n_{i}}+2Q_{Rbi}^{*}m_{l_{b}}^{2}\right)C_{2}\right]\\
A_{R}\left(n_{i}H_{R}^{+}W_{2}^{-}\right)= & \frac{i\sqrt{2}\left(\alpha_{13}k_{1}^{2}+2\rho_{1}v_{R}^{2}\right)g^{2}m_{l_{a}}}{128\pi^{2}\rho_{1}k_{1}v_{R}^{2}}\left(-Q_{Rai}K_{bi}^{*}m_{n_{i}}C_{0}-Q_{Rai}Q_{Rbi}^{*}m_{l_{b}}^{2}C_{2}-Q_{Rai}\left(K_{bi}^{*}m_{n_{i}}-2Q_{Rbi}^{*}m_{l_{b}}^{2}\right)C_{1}\right)
\end{align*}

\subsection{Two fermions in the loop}

For triangles $Xn_{i}n_{j}$ with $X=S,V$ and $S=G_{L}^{\pm},G_{R}^{\pm},H_{R}^{\pm}$,
$V=W,W^{\prime}$, the PV functions are given by

\begin{align*}
C_{0,1,2}= & C_{0}(m_{H_{1}^{0}},m_{l_{a}},m_{l_{b}},m_{X},m_{n_{i}},m_{n_{j}})\\
B_{0}^{\left(12\right)}= & B_{0}^{\left(12\right)}(m_{X},m_{n_{i}},m_{n_{j}})
\end{align*}
with 
\begin{align*}
\eta_{ij}^{ab}= & Q_{Lai}T_{RLjb}m_{W}^{2}C_{0}+Q_{Lai}T_{RLjb}B_{0}^{\left(12\right)}+Q_{Lai}m_{l_{b}}^{2}\left(T_{RLjb}-Q_{Lbj}^{*}m_{n_{j}}\right)C_{2}+T_{RLjb}\left(-Q_{Lai}m_{l_{a}}^{2}+T_{RLia}^{*}m_{n_{i}}\right)C_{1}\\
\omega_{ij}^{ab}= & Q_{Lbj}^{*}m_{l_{b}}^{2}\left(-Q_{Lai}m_{n_{i}}+T_{RLia}^{*}\right)C_{2}+T_{RLia}^{*}\left(T_{RLjb}m_{n_{j}}-Q_{Lbj}^{*}m_{l_{b}}^{2}\right)C_{1}\\
 & +\left(Q_{Lai}T_{RLjb}m_{n_{i}}m_{n_{j}}-Q_{Lai}Q_{Lbj}^{*}m_{l_{b}}^{2}m_{n_{i}}-T_{RLjb}T_{RLia}^{*}m_{n_{j}}+Q_{Lbj}^{*}T_{RLia}^{*}m_{l_{b}}^{2}\right)C_{0}\\
\gamma_{ij}^{ab}= & Q_{Lbj}^{*}T_{RLia}^{*}m_{W}^{2}C_{0}+Q_{Lbj}^{*}T_{RLia}^{*}B_{0}^{\left(12\right)}+Q_{Lbj}^{*}m_{l_{a}}^{2}\left(Q_{Lai}m_{n_{i}}-T_{RLia}^{*}\right)C_{1}+T_{RLia}^{*}\left(-T_{RLjb}m_{n_{j}}+Q_{Lbj}^{*}m_{l_{b}}^{2}\right)C_{2}\\
\delta_{ij}^{ab}= & Q_{Lai}m_{l_{a}}^{2}\left(-T_{RLjb}+Q_{Lbj}^{*}m_{n_{j}}\right)C_{1}+T_{RLjb}\left(Q_{Lai}m_{l_{a}}^{2}-T_{RLia}^{*}m_{n_{i}}\right)C_{2}\\
 & +\left(Q_{Lai}T_{RLjb}m_{l_{a}}^{2}-Q_{Lai}Q_{Lbj}^{*}m_{l_{a}}^{2}m_{n_{j}}-T_{RLjb}T_{RLia}^{*}m_{n_{i}}+Q_{Lbj}^{*}T_{RLia}^{*}m_{n_{i}}m_{n_{j}}\right)C_{0}\\
\kappa_{ij}^{ab}= & J_{ai}Q_{Rbj}^{*}m_{W^{\prime}}^{2}C_{0}+J_{ai}Q_{Rbj}^{*}B_{0}^{\left(12\right)}+J_{ai}\left(-J_{bj}^{*}m_{n_{j}}+Q_{Rbj}^{*}m_{l_{b}}^{2}\right)C_{2}+Q_{Rbj}^{*}m_{l_{a}}^{2}\left(-J_{ai}+Q_{Rai}m_{n_{i}}\right)C_{1}\\
\xi_{ij}^{ab}= & Q_{Rai}m_{l_{a}}^{2}\left(-J_{bj}^{*}+Q_{Rbj}^{*}m_{n_{j}}\right)C_{1}+J_{bj}^{*}\left(-J_{ai}m_{n_{i}}+Q_{Rai}m_{l_{a}}^{2}\right)C_{2}\\
 & +\left(-J_{ai}J_{bj}^{*}m_{n_{i}}+J_{ai}Q_{Rbj}^{*}m_{n_{i}}m_{n_{j}}+Q_{Rai}J_{bj}^{*}m_{l_{a}}^{2}-Q_{Rai}Q_{Rbj}^{*}m_{l_{a}}^{2}m_{n_{j}}\right)C_{0}\\
\varrho_{ij}^{ab}= & Q_{Rai}J_{bj}^{*}m_{W^{\prime}}^{2}C_{0}+Q_{Rai}J_{bj}^{*}B_{0}^{\left(12\right)}+Q_{Rai}m_{l_{b}}^{2}\left(J_{bj}^{*}-Q_{Rbj}^{*}m_{n_{j}}\right)C_{2}+J_{bj}^{*}\left(J_{ai}m_{n_{i}}-Q_{Rai}m_{l_{a}}^{2}\right)C_{1}\\
\sigma_{ij}^{ab}= & J_{ai}\left(J_{bj}^{*}m_{n_{j}}-Q_{Rbj}^{*}m_{l_{b}}^{2}\right)C_{1}+Q_{Rbj}^{*}m_{l_{b}}^{2}\left(J_{ai}-Q_{Rai}m_{n_{i}}\right)C_{2}\\
 & +\left(-J_{ai}J_{bj}^{*}m_{n_{j}}+J_{ai}Q_{Rbj}^{*}m_{l_{b}}^{2}+Q_{Rai}J_{bj}^{*}m_{n_{i}}m_{n_{j}}-Q_{Rai}Q_{Rbj}^{*}m_{l_{b}}^{2}m_{n_{i}}\right)C_{0}\\
\vartheta_{ij}^{ab}= & K_{ai}Q_{Rbj}^{*}\left(m_{H_{R}^{+}}\right)^{2}C_{0}+K_{ai}Q_{Rbj}^{*}B_{0}^{\left(12\right)}+K_{ai}\left(-K_{bj}^{*}m_{n_{j}}+Q_{Rbj}^{*}m_{l_{b}}^{2}\right)C_{2}+Q_{Rbj}^{*}m_{l_{a}}^{2}\left(-K_{ai}+Q_{Rai}m_{n_{i}}\right)C_{1}\\
\varsigma_{ij}^{ab}= & Q_{Rai}m_{l_{a}}^{2}\left(-K_{bj}^{*}+Q_{Rbj}^{*}m_{n_{j}}\right)C_{1}+K_{bj}^{*}\left(-K_{ai}m_{n_{i}}+Q_{Rai}m_{l_{a}}^{2}\right)C_{2}\\
 & +\left(-K_{ai}K_{bj}^{*}m_{n_{i}}+K_{ai}Q_{Rbj}^{*}m_{n_{i}}m_{n_{j}}+Q_{Rai}K_{bj}^{*}m_{l_{a}}^{2}-Q_{Rai}Q_{Rbj}^{*}m_{l_{a}}^{2}m_{n_{j}}\right)C_{0}\\
\varphi_{ij}^{ab}= & Q_{Rai}K_{bj}^{*}\left(m_{H_{R}^{+}}\right)^{2}C_{0}+Q_{Rai}K_{bj}^{*}B_{0}^{\left(12\right)}+Q_{Rai}m_{l_{b}}^{2}\left(K_{bj}^{*}-Q_{Rbj}^{*}m_{n_{j}}\right)C_{2}+K_{bj}^{*}\left(K_{ai}m_{n_{i}}-Q_{Rai}m_{l_{a}}^{2}\right)C_{1}\\
\phi_{ij}^{ab}= & K_{ai}\left(K_{bj}^{*}m_{n_{j}}-Q_{Rbj}^{*}m_{l_{b}}^{2}\right)C_{1}+Q_{Rbj}^{*}m_{l_{b}}^{2}\left(K_{ai}-Q_{Rai}m_{n_{i}}\right)C_{2}\\
 & +\left(-K_{ai}K_{bj}^{*}m_{n_{j}}+K_{ai}Q_{Rbj}^{*}m_{l_{b}}^{2}+Q_{Rai}K_{bj}^{*}m_{n_{i}}m_{n_{j}}-Q_{Rai}Q_{Rbj}^{*}m_{l_{b}}^{2}m_{n_{i}}\right)C_{0}
\end{align*}
the form factors are given by
\begin{align*}
A_{L}\left(G_{L}^{\pm}n_{i}n_{j}\right)= & \frac{i\sqrt{2}m_{l_{a}}}{16\pi^{2}k_{1}^{3}}\left[\left(\omega_{ij}^{ab}\Omega_{RLij}+\eta_{ij}^{ab}\Omega_{RLij}^{*}\right)-\frac{\alpha_{13}\epsilon^{2}}{2\rho_{1}}\left(\eta_{ij}^{ab}\Omega_{SRij}+\omega_{ij}^{ab}\Omega_{SRij}^{*}\right)\right]\\
A_{R}\left(G_{L}^{\pm}n_{i}n_{j}\right)= & \frac{i\sqrt{2}m_{l_{b}}}{16\pi^{2}k_{1}^{3}}\left[\left(\gamma_{ij}^{ab}\Omega_{RLij}+\delta_{ij}^{ab}\Omega_{RLij}^{*}\right)-\frac{\alpha_{13}\epsilon^{2}}{2\rho_{1}}\left(\delta_{ij}^{ab}\Omega_{SRij}+\gamma_{ij}^{ab}\Omega_{SRij}^{*}\right)\right]\\
A_{L}\left(G_{R}^{\pm}n_{i}n_{j}\right)= & \frac{i\sqrt{2}m_{l_{b}}}{16\pi^{2}k_{1}v_{R}^{2}}\left[\left(\xi_{ij}^{ab}\Omega_{RLij}+\kappa_{ij}^{ab}\Omega_{RLij}^{*}\right)-\frac{\alpha_{13}\epsilon^{2}}{2\rho_{1}}\left(\kappa_{ij}^{ab}\Omega_{SRij}+\xi_{ij}^{ab}\Omega_{SRij}^{*}\right)\right]\\
A_{R}\left(G_{R}^{\pm}n_{i}n_{j}\right)= & \frac{i\sqrt{2}m_{l_{a}}}{16\pi^{2}k_{1}v_{R}^{2}}\left[\left(\varrho_{ij}^{ab}\Omega_{RLij}+\sigma_{ij}^{ab}\Omega_{RLij}^{*}\right)-\frac{\alpha_{13}\epsilon^{2}}{2\rho_{1}}\left(\sigma_{ij}^{ab}\Omega_{SRij}+\varrho_{ij}^{ab}\Omega_{SRij}^{*}\right)\right]\\
A_{L}\left(H_{R}^{\pm}n_{i}n_{j}\right)= & \frac{i\sqrt{2}m_{l_{b}}}{16\pi^{2}k_{1}^{3}}\left[\left(\varsigma_{ij}^{ab}\Omega_{RLij}+\vartheta_{ij}^{ab}\Omega_{RLij}^{*}\right)-\frac{\alpha_{13}\epsilon^{2}}{2\rho_{1}}\left(\vartheta_{ij}^{ab}\Omega_{SRij}+\varsigma_{ij}^{ab}\Omega_{SRij}^{*}\right)\right]\\
A_{R}\left(H_{R}^{\pm}n_{i}n_{j}\right)= & \frac{i\sqrt{2}m_{l_{a}}}{16\pi^{2}k_{1}^{3}}\left[\left(\varphi_{ij}^{ab}\Omega_{RLij}+\phi_{ij}^{ab}\Omega_{RLij}^{*}\right)-\frac{\alpha_{13}\epsilon^{2}}{2\rho_{1}}\left(\phi_{ij}^{ab}\Omega_{SRij}+\varphi_{ij}^{ab}\Omega_{SRij}^{*}\right)\right]
\end{align*}
\begin{align*}
A_{L}\left(W^{\pm}n_{i}n_{j}\right)= & \frac{\sqrt{2}ig^{2}m_{l_{a}}}{64\pi^{2}k_{1}}\left[-\Omega_{RLij}m_{n_{i}}C_{1}+\Omega_{RLij}^{*}\left(C_{0}-C_{1}\right)m_{n_{j}}\right.\\
 & \left.+\frac{\alpha_{13}\epsilon^{2}}{2\rho_{1}}\left(\Omega_{SRij}^{*}m_{n_{i}}C_{1}-\Omega_{SRij}\left(C_{0}-C_{1}\right)m_{n_{j}}\right)\right]Q_{Lai}Q_{Lbj}^{*}\\
A_{R}\left(W^{\pm}n_{i}n_{j}\right)= & \frac{\sqrt{2}ig^{2}m_{l_{b}}}{64\pi^{2}k_{1}}\left[\Omega_{RLij}^{*}m_{n_{j}}C_{2}+\Omega_{RLij}\left(C_{0}+C_{2}\right)m_{n_{i}}\right.\\
 & \left.-\frac{\alpha_{13}\epsilon^{2}}{2\rho_{1}}\left(\Omega_{SRij}m_{n_{j}}C_{2}+\Omega_{SRij}^{*}\left(C_{0}+C_{2}\right)m_{n_{i}}\right)\right]Q_{Lai}Q_{Lbj}^{*}\\
A_{L}\left(W^{\prime\pm}n_{i}n_{j}\right)= & \frac{\sqrt{2}ig^{2}m_{l_{b}}}{64\pi^{2}k_{1}}\left[\Omega_{RLij}m_{n_{j}}C_{2}+\Omega_{RLij}^{*}\left(C_{0}+C_{2}\right)m_{n_{i}}\right.\\
 & \left.-\frac{\alpha_{13}\epsilon^{2}}{2\rho_{1}}\left(\Omega_{SRij}^{*}m_{n_{j}}C_{2}+\Omega_{SRij}\left(C_{0}+C_{2}\right)m_{n_{i}}\right)\right]Q_{Rai}Q_{Rbj}^{*}\\
A_{R}\left(W^{\prime\pm}n_{i}n_{j}\right)= & \frac{\sqrt{2}ig^{2}m_{l_{a}}}{64\pi^{2}k_{1}}\left[-\Omega_{RLij}^{*}m_{n_{i}}C_{1}+\Omega_{RLij}\left(C_{0}-C_{1}\right)m_{n_{j}}\right.\\
 & \left.+\frac{\alpha_{13}\epsilon^{2}}{2\rho_{1}}\left(\Omega_{SRij}m_{n_{i}}C_{1}-\Omega_{SRij}^{*}\left(C_{0}-C_{1}\right)m_{n_{j}}\right)\right]Q_{Rai}Q_{Rbj}^{*}
\end{align*}

\bibliographystyle{unsrt}
\bibliography{biblioteca}

\begin{thebibliography}{10}

\bibitem{Pati:1974yy}
Jogesh~C. Pati and Abdus Salam.
\newblock Lepton number as the fourth "color".
\newblock {\em Phys. Rev. D}, 10:275--289, Jul 1974.

\bibitem{Mohapatra:1974gc}
R.~N. Mohapatra and J.~C. Pati.
\newblock "natural" left-right symmetry.
\newblock {\em Phys. Rev. D}, 11:2558--2561, May 1975.

\bibitem{Mohapatra:1974hk}
Rabindra~N. Mohapatra and Jogesh~C. Pati.
\newblock Left-right gauge symmetry and an "isoconjugate" model of
  $\mathrm{CP}$ violation.
\newblock {\em Phys. Rev. D}, 11:566--571, Feb 1975.

\bibitem{Senjanovic:1975rk}
G.~Senjanovic and R.~N. Mohapatra.
\newblock Exact left-right symmetry and spontaneous violation of parity.
\newblock {\em Phys. Rev. D}, 12:1502--1505, Sep 1975.

\bibitem{SENJANOVIC1979334}
Goran Senjanović.
\newblock Spontaneous breakdown of parity in a class of gauge theories.
\newblock {\em Nuclear Physics B}, 153:334--364, 1979.

\bibitem{Dev:2016dja}
P.~S. Bhupal~Dev, Rabindra~N. Mohapatra, and Yongchao Zhang.
\newblock Displaced photon signal from a possible light scalar in minimal
  left-right seesaw model.
\newblock {\em Phys. Rev. D}, 95:115001, Jun 2017.

\bibitem{Bertolini:2014sua}
Stefano Bertolini, Alessio Maiezza, and Fabrizio Nesti.
\newblock {Present and Future K and B Meson Mixing Constraints on TeV Scale
  Left-Right Symmetry}.
\newblock {\em Phys. Rev. D}, 89(9):095028, 2014.

\bibitem{Mohapatra:1986bd}
Rabindra~N. Mohapatra.
\newblock Mechanism for understanding small neutrino mass in superstring
  theories.
\newblock {\em Phys. Rev. Lett.}, 56:561--563, Feb 1986.

\bibitem{Dev:2012sg}
P.~S.~Bhupal Dev and R.~N. Mohapatra.
\newblock {TeV Scale Inverse Seesaw in SO(10) and Leptonic Non-Unitarity
  Effects}.
\newblock {\em Phys. Rev. D}, 81:013001, 2010.

\bibitem{Cheng:1976uq}
Ta-Pei Cheng and Ling-Fong Li.
\newblock Muon-number-nonconservation effects in a gauge theory with $v + a$
  currents and heavy neutral leptons.
\newblock {\em Phys. Rev. D}, 16:1425--1443, Sep 1977.

\bibitem{PhysRevD.71.035011}
Ernesto Arganda, Ana~M. Curiel, Mar\'{\i}a~J. Herrero, and David Temes.
\newblock Lepton flavor violating higgs boson decays from massive seesaw
  neutrinos.
\newblock {\em Phys. Rev. D}, 71:035011, Feb 2005.

\bibitem{PhysRevD.91.015001}
E.~Arganda, M.~J. Herrero, X.~Marcano, and C.~Weiland.
\newblock Imprints of massive inverse seesaw model neutrinos in lepton flavor
  violating higgs boson decays.
\newblock {\em Phys. Rev. D}, 91:015001, Jan 2015.

\bibitem{THAO2017159}
N.H. Thao, L.T. Hue, H.T. Hung, and N.T. Xuan.
\newblock Lepton flavor violating higgs boson decays in seesaw models: New
  discussions.
\newblock {\em Nuclear Physics B}, 921:159--180, 2017.

\bibitem{HUE201637}
L.T. Hue, H.N. Long, T.T. Thuc, and T.~{Phong Nguyen}.
\newblock Lepton flavor violating decays of standard-model-like higgs in 3-3-1
  model with neutral lepton.
\newblock {\em Nuclear Physics B}, 907:37--76, 2016.

\bibitem{Vicente2019HiggsLF}
Avelino Vicente.
\newblock {Higgs lepton flavor violating decays in Two Higgs Doublet Models}.
\newblock {\em Front. in Phys.}, 7:174, 2019.

\bibitem{Arroyo-Urena:2023vfh}
M.~A. Arroyo-Ure{\~n}a, J.~Lorenzo D{\'\i}az-Cruz, O.~F{\'e}lix-Beltr{\'a}n,
  and M.~Zeleny-Mora.
\newblock {Lessons from LHC on the LFV Higgs decays
  h{\textrightarrow}{\ensuremath{\ell}}a{\ensuremath{\ell}}b in the two-Higgs
  doublet models}.
\newblock {\em Int. J. Mod. Phys. A}, 39(21):2450079, 2024.

\bibitem{PhysRevD.108.036002}
Fang Xu.
\newblock Neutral and doubly charged scalars at future lepton colliders.
\newblock {\em Phys. Rev. D}, 108:036002, Aug 2023.

\bibitem{Asadi:2025dii}
Pouya Asadi, Hengameh Bagherian, Katherine Fraser, Samuel Homiller, and Qianshu
  Lu.
\newblock {Lepton Flavor Violation: From Muon Decays to Muon Colliders}.
\newblock 9 2025.

\bibitem{MEGII:2025gzr}
K.~Afanaciev et~al.
\newblock {New limit on the
  {\ensuremath{\mu}}+-{\ensuremath{>}}e+{\ensuremath{\gamma}}decay with the MEG
  II experiment}.
\newblock 4 2025.

\bibitem{BaBar:2009hkt}
Bernard Aubert et~al.
\newblock {Searches for Lepton Flavor Violation in the Decays $\tau^\pm \to
  e^\pm \gamma$ and $\tau^\pm \to \mu^\pm \gamma$}.
\newblock {\em Phys. Rev. Lett.}, 104:021802, 2010.

\bibitem{Belle-II:2018jsg}
W.~Altmannshofer et~al.
\newblock {The Belle II Physics Book}.
\newblock {\em PTEP}, 2019(12):123C01, 2019.
\newblock [Erratum: PTEP 2020, 029201 (2020)].

\bibitem{Belle:2021ysv}
A.~Abdesselam et~al.
\newblock {Search for lepton-flavor-violating tau-lepton decays to $\ell\gamma$
  at Belle}.
\newblock {\em JHEP}, 10:19, 2021.

\bibitem{ATLAS:2023mvd}
{Searches for lepton-flavour-violating decays of the Higgs boson into $e\tau$
  and $\mu\tau$ in $\sqrt{s}=13$ TeV $pp$ collisions with the ATLAS detector}.
\newblock 2 2023.

\bibitem{CMS:2023pte}
Aram Hayrapetyan et~al.
\newblock {Search for the lepton-flavor violating decay of the Higgs boson and
  additional Higgs bosons in the e$\mu$ final state in proton-proton collisions
  at $\sqrt{s}$ = 13 TeV}.
\newblock {\em Phys. Rev. D}, 108(7):072004, 2023.

\bibitem{Gu:2010xc}
Pei-Hong Gu and Utpal Sarkar.
\newblock {Leptogenesis with Linear, Inverse or Double Seesaw}.
\newblock {\em Phys. Lett. B}, 694:226--232, 2011.

\bibitem{Brdar:2018sbk}
Vedran Brdar and Alexei~Yu Smirnov.
\newblock {Low Scale Left-Right Symmetry and Naturally Small Neutrino Mass}.
\newblock {\em JHEP}, 02:045, 2019.

\bibitem{Catano:2012kw}
M.~E. Catano, R~Martinez, and F.~Ochoa.
\newblock {Neutrino masses in a 331 model with right-handed neutrinos without
  doubly charged Higgs bosons via inverse and double seesaw mechanisms}.
\newblock {\em Phys. Rev. D}, 86:073015, 2012.

\bibitem{PhysRevD.86.035007}
A.~G. Dias, C.~A. de~S.~Pires, P.~S. Rodrigues~da Silva, and A.~Sampieri.
\newblock Simple realization of the inverse seesaw mechanism.
\newblock {\em Phys. Rev. D}, 86:035007, Aug 2012.

\bibitem{Casas:2001sr}
J.~A. Casas and A.~Ibarra.
\newblock {Oscillating neutrinos and $\mu \to e, \gamma$}.
\newblock {\em Nucl. Phys. B}, 618:171--204, 2001.

\bibitem{Zeleny-Mora:2021tym}
M.~Zeleny-Mora, J.~Lorenzo D\'\i{}az-Cruz, and O.~F\'elix-Beltr\'an.
\newblock {The general one-loop structure for the LFV Higgs decays
  Hr\textrightarrow{}lalb in multi-Higgs models with neutrino masses}.
\newblock {\em Int. J. Mod. Phys. A}, 37(36):2250226, 2022.

\bibitem{Esteban:2024eli}
Ivan Esteban, M.~C. Gonzalez-Garcia, Michele Maltoni, Ivan Martinez-Soler,
  Jo{\~a}o~Paulo Pinheiro, and Thomas Schwetz.
\newblock {NuFit-6.0: updated global analysis of three-flavor neutrino
  oscillations}.
\newblock {\em JHEP}, 12:216, 2024.

\bibitem{Staub:2011dp}
Florian Staub, Thorsten Ohl, Werner Porod, and Christian Speckner.
\newblock {A Tool Box for Implementing Supersymmetric Models}.
\newblock {\em Comput. Phys. Commun.}, 183:2165--2206, 2012.

\bibitem{Staub:2015kfa}
Florian Staub.
\newblock {Exploring new models in all detail with SARAH}.
\newblock {\em Adv. High Energy Phys.}, 2015:840780, 2015.

\bibitem{Porod:2011nf}
W.~Porod and F.~Staub.
\newblock {SPheno 3.1: Extensions including flavour, CP-phases and models
  beyond the MSSM}.
\newblock {\em Comput. Phys. Commun.}, 183:2458--2469, 2012.

\bibitem{BREIN200542}
Oliver Brein.
\newblock Adaptive scanning—a proposal how to scan theoretical predictions
  over a multi-dimensional parameter space efficiently.
\newblock {\em Computer Physics Communications}, 170(1):42--48, 2005.

\bibitem{Diaz:2024sxg}
Mauricio~A. Diaz, Srinandan Dasmahapatra, and Stefano Moretti.
\newblock {hep-aid: A Python Library for Sample Efficient Parameter Scans in
  Beyond the Standard Model Phenomenology}.
\newblock 12 2024.

\bibitem{Diaz:2024yfu}
Mauricio~A. Diaz, Giorgio Cerro, Srinandan Dasmahapatra, and Stefano Moretti.
\newblock {Bayesian Active Search on Parameter Space: a 95 GeV Spin-0 Resonance
  in the ($B-L$)SSM}.
\newblock 4 2024.

\bibitem{Hahn:1998yk}
T.~Hahn and M.~Perez-Victoria.
\newblock {Automatized one loop calculations in four-dimensions and
  D-dimensions}.
\newblock {\em Comput. Phys. Commun.}, 118:153--165, 1999.

\bibitem{Lichtenstein:2025pxs}
Gabriela Lichtenstein, Ricardo~C. Silva, Mario~J. Neves, and Farinaldo Queiroz.
\newblock {Updated Bounds on the Minimal Left-Right Symmetric Model from LHC
  Dilepton Resonance Searches}.
\newblock 7 2025.

\bibitem{ThomasArun:2021rwf}
Mathew Thomas~Arun, Tanumoy Mandal, Subhadip Mitra, Ananya Mukherjee, Lakshmi
  Priya, and Adithya Sampath.
\newblock {Testing left-right symmetry with an inverse seesaw mechanism at the
  LHC}.
\newblock {\em Phys. Rev. D}, 105(11):115007, 2022.

\end{thebibliography}

\end{document}